\documentclass[useAMS,usenatbib]{mn2e}

\usepackage{graphicx}

\title[CO(1--0) in star-forming galaxies at $z\sim2$]{CO(1--0) line imaging of massive star-forming disk galaxies at $z=1.5-2.2$}
\author[M.~Aravena et al.]
{\parbox{\textwidth}{M.~Aravena$^{1,2}$\thanks{E-mail: manuel.aravenaa@mail.udp.cl},~J.~A.~Hodge$^{3}$,~J.~Wagg$^{4}$,~C.~L.~Carilli$^{3,5}$,~E.~Daddi$^{6}$,~H.~Dannerbauer$^{7}$, L.~Lentati$^{5}$,~D.~A.~Riechers$^{8}$,~M.~Sargent$^{9}$~and~F.~Walter$^{10,3}$
}\vspace{0.5cm}\\
\parbox{\textwidth}{
$^{1}$European Southern Observatory, Alonso de C\'{o}rdoba 3107, Casilla 19001, Vitacura, Santiago, Chile\\
$^{2}$N\'{u}cleo de Astronom\'{\i}a, Facultad de Ingenier\'{\i}a, Universidad Diego Portales, Av. Ej\'{e}rcito 441, Santiago, Chile\\
$^{3}$National Radio Astronomy Observatory, 520 Edgemont Rd., Charlottesville VA, USA \\
$^{4}$Square Kilometer Array organization, Lower Withington, UK\\
$^{5}$Cavendish Laboratory, University of Cambridge, J. J. Thompson Ave., Cambridge CB3 0HE, UK\\
$^{6}$CEA, IRFU/SAp, F-91191, Gif-Sur-Yvette, France\\
$^{7}$Universit\"{a}t Wien, Institut f\"{u}r Astrophysik, T\"{u}rkenstra{\ss}e 17, 1180 Wien, Austria\\
$^{8}$Department of Astronomy, Space Science Building, Cornell University, Ithaca, NY 14853-6801, USA\\
$^{9}$Astronomy Centre, Dept. of Physics and Astronomy, University of Sussex, Falmer, Brighton BN1 9QH, UK\\
$^{10}$Max-Planck Institute for Astronomy, K\"{o}nisgstuhl 17, D-69117, Heidelberg, Germany\\
}}
\begin{document}

%\date{January 10, 2012}

\pagerange{\pageref{firstpage}--\pageref{lastpage}} \pubyear{2014}

\maketitle
\label{firstpage}
%\clearpage
\begin{abstract}
We present detections of the CO($J=$ 1--0) emission line in a sample of four massive star-forming galaxies at $z\sim1.5-2.2$ obtained with the Karl G. Jansky Very Large Array (VLA). Combining these observations with previous CO(2--1) and CO(3--2) detections of these galaxies, we study the excitation properties of the molecular gas in our sample sources. We find an average line brightness temperature ratios of $R_{21}=0.70\pm0.16$ and $R_{31}=0.50\pm0.29$, based on measurements for three and two galaxies, respectively. These results provide additional support to previous indications of sub-thermal gas excitation for the CO(3--2) line with a typically assumed line ratio $R_{31}\sim0.5$. For one of our targets, BzK-21000, we present spatially resolved CO line maps. At the resolution of $0.18''$ (1.5 kpc), most of the emission is resolved out except for some clumpy structure. From this, we attempt to identify molecular gas clumps in the data cube, finding 4 possible candidates. We estimate that $<40\%$ of the molecular gas is confined to giant clumps ($\sim1.5$ kpc in size), and thus most of the gas could be distributed in small fainter clouds or in fairly diffuse extended regions of lower brightness temperatures than our sensitivity limit. 
%Finally, we argue that finding such clumps using high-resolution CO imaging in massive disk galaxies at $z>1$, or fainter objects, is at the very limit of what can be done with VLA, and future studies will need to use the Atacama Large Millimeter/submillimeter Array (ALMA), albeit at higher-J transitions.
\end{abstract}


\section{Introduction}
The physical processes behind gas supply, and subsequent star formation, in early galaxies, remain key uncertainties in our understanding of galaxy formation. Study of the demographics and gas depletion timescales show that most star forming galaxies at $z>1$ form a tight correlation between their star formation rates (SFR) and stellar masses. Such correlation, usually termed as the `main-sequence'  of galaxies, suggests that the star formation is typically a long-lived process, likely occurring over timescales of $\sim0.7$ Gyr \citep[e.g., ][]{Daddi2005,Daddi2007}. Major gas rich mergers, typically lying above such main-sequence, are unlikely to be the primary driving mechanism for such continuous star formation, since they would lead to short timescale, nuclear starbursts as it is seen in local Ultra Luminous IR Galaxies \citep[ULIRGs; e.g., ][]{Solomon1997, Downes1998}.

%The so-called cold-mode accretion model has been proposed as the dominant mechanism to feed gas to galaxies. Critical tests of the merger-driven and cold mode accretion models come from the study of the gas dynamics and distribution \citep{Keres2005, Dekel2009a, Dekel2009b}. In mergers, an intense burst of star formation is triggered by rapid gas inflows toward the inner central kpc. The gas distribution appears concentrated in a central sub-kpc region with high gas and star-formation rate (SFR) surface densities, possibly showing signs of interactions (tidal features). The gas is depleted in $<100$ Myr, and the velocity field appears chaotic, without dominant rotation. Conversely, in the cold mode accretion model the gas continuously streams from the inter-galactic medium along cool, dense filaments forming a thick, rotating disk that is punctuated by several giant clumps of enhanced gas and SFR densities on scales $\sim1$ kpc in clump size. 
Two of the major challenges for understanding the mechanisms of gas accretion and stellar build up in these galaxies is to spatially and kinematically resolve the star formation and molecular gas components down to kpc scales \citep{Shapiro2008}, as well as targeting several molecular gas emission lines in order to characterize the physical state of the interstellar medium \citep[ISM;][]{Carilli2013}. High-resolution observations of the ionized gas (H$\alpha$) kinematics and the stellar component have shown that a majority of these galaxies are consistent with clumpy rotating disks with sizes of order 10 kpc \citep{Genzel2006, Genzel2008, ForsterSchreiber2009, ForsterSchreiber2011}. %, with the most massive galaxies showing powerful gas outflows which exceed the SFR of the host galaxies \citep{Shapiro2009, Genzel2011}. 
However, these observations do not directly trace the cold molecular gas from which stars are formed, and could be prone to obscuration within the host galaxy. Direct high resolution observations of the molecular gas are thus necessary.

\begin{table*}
\centering
\caption{Observational parameters\label{tab:0}}
\begin{tabular}{lcccccccc}
\hline
Source & $\alpha_{\rm CO(1-0)}$ $^{\rm a}$& $\delta_{\rm CO(1-0)}$ $^{\rm a}$& Array $^{\rm b}$&  $\nu_{\rm obs}$ $^{\rm c}$& Beam $^{\rm d}$& P.A. $^{\rm d}$& rms $^{\rm e}$& Chan $^{\rm f}$\\ 
               &  (J2000)                   & (J2000)                   &                    &  (GHz)                  &            &  (deg) & ($\mu$Jy) & (km s$^{-1}$)\\
\hline\hline
BzK-21000 &  $12^{\rm h}37^{\rm m}10.69^{\rm  s}$   & $+62^\circ22'34.3''$ &  B/D  &  45.656 &  $0.18''\times0.17''$ & $-178$ & 91 & 50 \\
BzK-4171&  $12^{\rm h}36^{\rm m}26.61^{\rm  s}$   & $+62^\circ08'35.6''$ &  D    &  46.765 &  $2.50''\times2.25''$ & $+47$ & 220 & 80 \\
BzK-16000 &  $12^{\rm h}36^{\rm m}30.09^{\rm  s}$   & $+62^\circ14'27.7''$ &  C/D    &  45.708 &  $1.46''\times1.31''$ & $-46$ & 130 & 80 \\
BX610 &  $23^{\rm h}46^{\rm m}09.45^{\rm  s}$   & $+12^\circ49'19.3''$ &  D    &  35.905 &  $1.65''\times1.46''$ & $+28$ & 180 & 80 \\
\hline
\end{tabular}
\begin{flushleft}
\begin{footnotesize}
\noindent {\bf Notes:} $^{\rm a}$ Right ascension and declination of the measured CO(1--0) positions; $^{\rm b}$ VLA array configuration used in the observations; $^{\rm c}$ Observed frequency; $^{\rm d} $Synthesized beam size (major and minor axis) and position angle (P.A.); $^{\rm e}$ Average noise level reached in the data cube at a velocity resolution per channel given in $^{\rm f}$.
\end{footnotesize}
\end{flushleft}
\end{table*}

Recently, observations of the molecular gas, through CO line emission, in optical/IR color-selected star-forming galaxies at $z=1-3$ have revealed significant molecular gas reservoirs, comparable to their stellar component ($\sim10^{10}$ M$_\odot$) in systems which typically show SFRs in the range $50-400$ M$_\odot$ yr$^{-1}$ \citep{Daddi2008, Daddi2010a, Tacconi2010, Tacconi2013}. This indicated that these galaxies have low star formation efficiencies and long gas depletion timescales, compared to that seen in extreme starbursting submillimeter galaxies (SMGs) and quasars, and suggest their integrated properties may follow a different star-formation law \citep{Daddi2010b, Genzel2010}.  Furthermore, their implied CO luminosity to gas mass conversion factors resemble local disk galaxies \citep{Daddi2010a, Magdis2011, Magdis2012,Magnelli2012}. 
%The ratio of their molecular to stellar mass is $\ge1$, as compared to typical values $< 0.1$ in the nearby Universe, even for gas rich galaxies such as the Milky Way \citep{Geach2011, Tacconi2010, Tacconi2013, Tan2013}. 

These studies, however, focused on the observation of $J>1$ CO line emission. One of the major assumptions encountered when observing $J>1$ CO lines in main-sequence galaxies has been the the adoption of an ``average'' constant ratio between the $J>1$ CO line and CO(1--0). Determination of these ratios are required in order to convert the high-J CO line luminosities into CO(1--0) luminosities, for which the conversion factors of CO luminosity to gas mass have been calibrated \citep{Bolatto2013}. Thus, observations of the CO(1--0) line in statistical samples of star-forming galaxies at $z>1$ are necessary for calibrating such line ratios. 

As Bauermeister et al. point out, calibration of the line ratios is particularly important for the CO(3--2) line for two main reasons: (1) this line ($\nu_{\rm rest}=345.538$ GHz) is shifted to the 2mm and 3mm atmospheric windows at $z=1-3$, being directly accessible with the most powerful (sub)millimeter facilities that can easily access these kind of galaxies, namely the Atacama Large Millimeter/submillimeter Array (ALMA) and the Plateau de Bureau Interferometer (PdBI). Hence, observations of the CO(3--2) line will (and have) become routine and constitute the first direct attempt to characterize the molecular gas properties of these objects \citep[e.g., ][]{Tacconi2010,Tacconi2013}; (2) The cosmic time spanned by redshifts $z=1-3$, 6 Gyr, corresponds to the important period when most of the stars in the Universe where created and where most of the galaxies were assembled. 

To date, only a few main-sequence galaxies at cosmological distances ($z>0.1$) have observations in two or more CO lines \citep{Dannerbauer2009, Aravena2010b, Bauermeister2013}, and generally only a handful of these main-sequence galaxies have spatially resolved CO observations down to $<10$ kpc scales \citep{Tacconi2013, Genzel2013}. Follow-up CO(1--0) observations of three representative $z\sim1.5$ galaxies in the CO(1--0) line emission suggest that the molecular gas in these systems is already sub-thermally excited at the CO(3--2) transition similar to what is found in local disks \citep{Dannerbauer2009, Aravena2010b}, with typical line brightness temperature ratios between both lines of $\sim0.5$. This is also similar to what is found in SMGs \citep{Harris2010, Ivison2011, Bothwell2013}, but substantially different to high-redshift QSOs, which appear to have highly excited gas with line temperature ratios close to unity \citep[e.g.,]{Riechers2006,Weiss2007,Ao2008,Riechers2011c, Ivison2012}. Recent observations of $z\sim0.3$ disk galaxies support these findings, indicating that the molecular gas content, as traced by CO(1--0), is two times larger than expected from $J>3$ CO measurements, comparable to $z>2$ SMGs \citep{Papadopoulos2002, Harris2010, Ivison2011, Riechers2011b}. 
%As such, understanding the processes of star formation in galaxies at those redshifts is of ultimate importance. 

In this paper, we present  observations of the CO(1-0) emission line in four main-sequence star forming galaxies at $z\sim1.5-2.2$ obtained with the Karl G. Jansky Very Large Array (VLA). The angular resolution of our observations for one of these sources, $0.18''$, allows us to spatially resolve the distribution of the molecular gas. The new correlator system at the VLA permits us to expand the bandwidth and velocity resolution of the previous CO(1--0) detections in three of our targets, and to detect the CO emission in a new object at $z=2.2$. We adopt a standard $\Lambda$CDM cosmology throughout \citep{Komatsu2011}.

\begin{figure*}
%\centering
\includegraphics[scale=0.35]{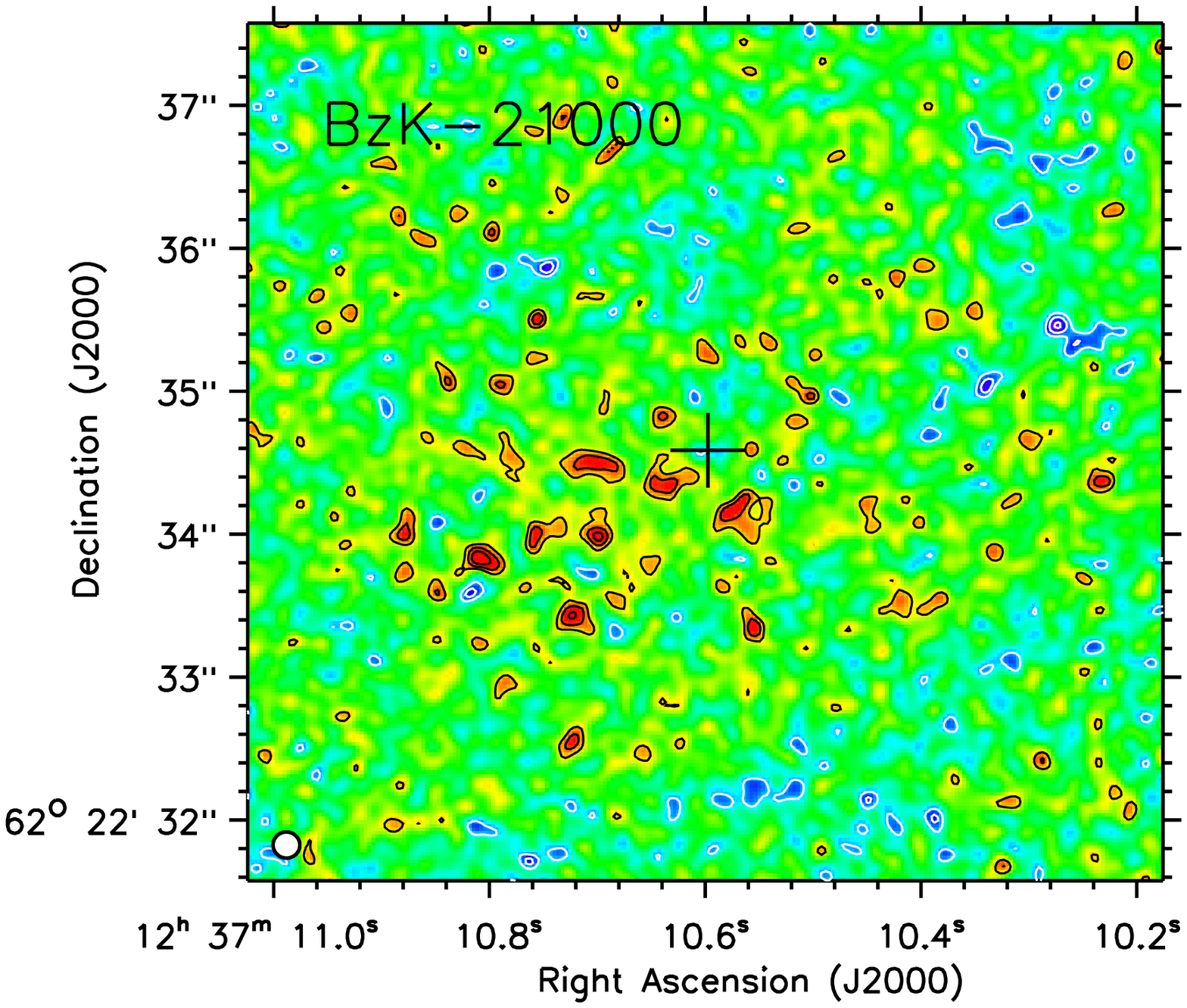}%\hspace{4mm}
\includegraphics[scale=0.35]{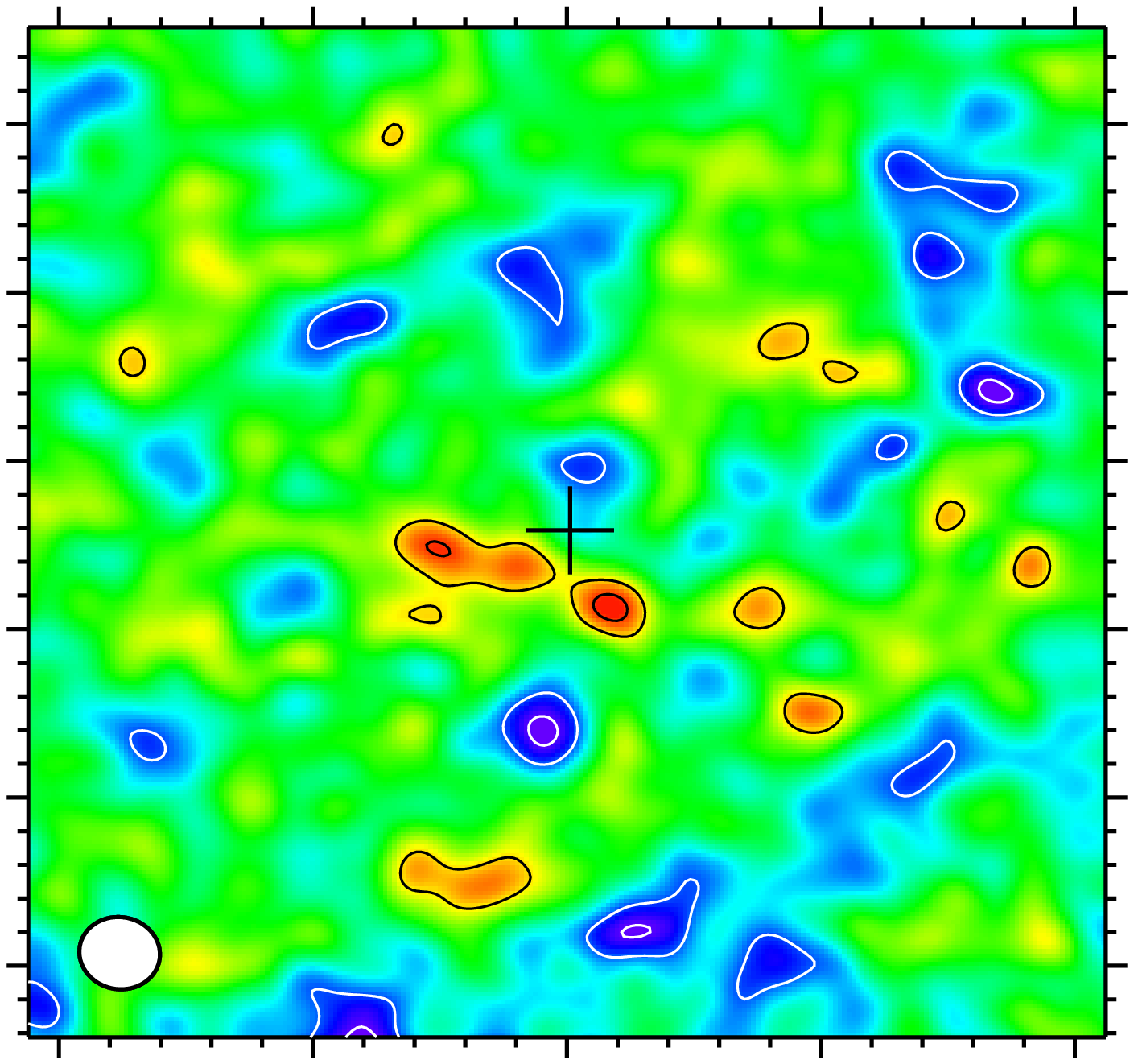}%\hspace{4mm}
\includegraphics[scale=0.35]{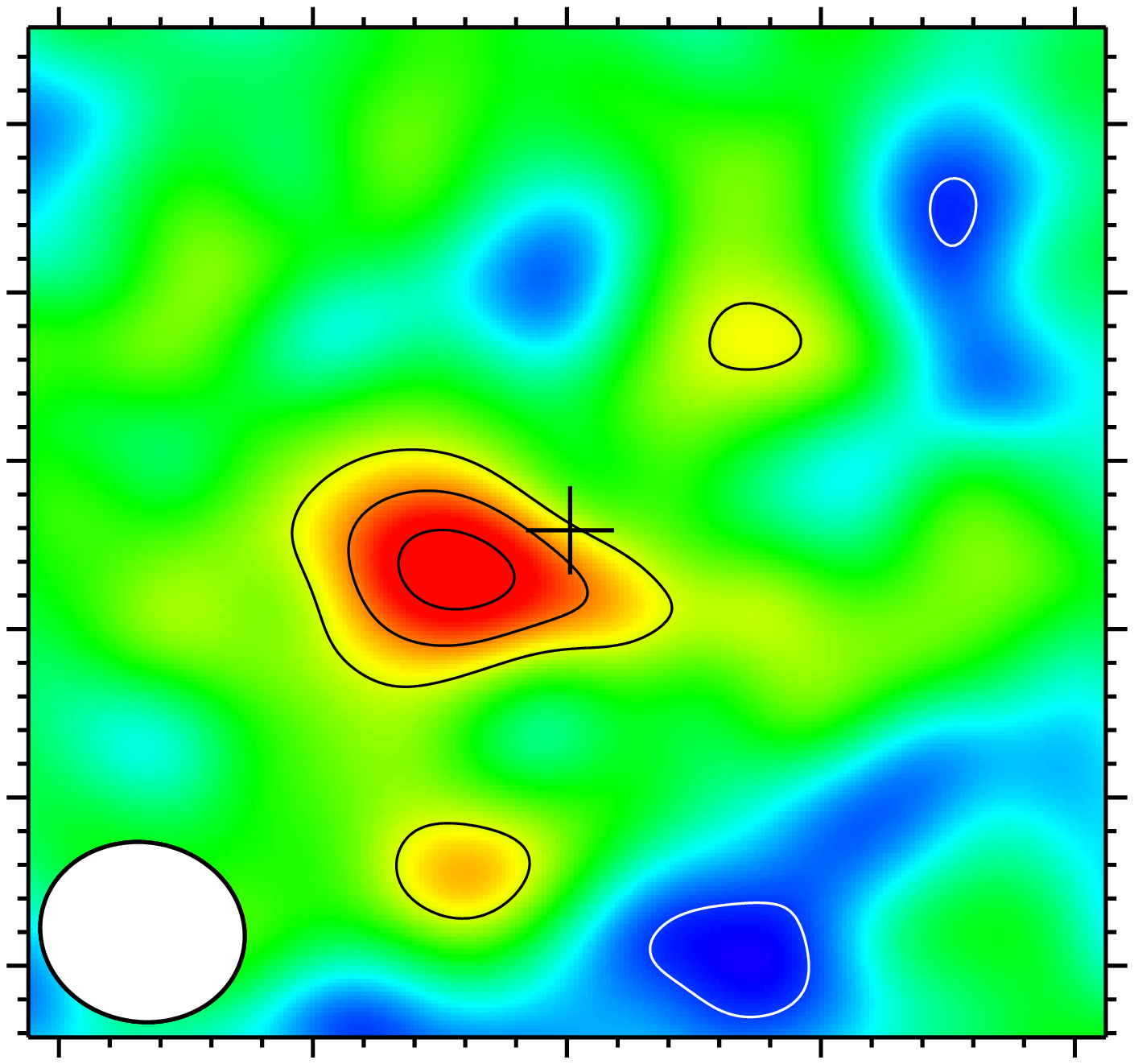}%\hspace{4mm}
\caption{VLA CO(1--0) maps averaged over 440 km s$^{-1}$ for BzK-21000, shown at the native resolution of $\sim0.18''$ (left), tapered to $0.4''$ (middle), and tapered to $1.1''$ (right). Contour levels are given in steps of $\pm1\sigma$, starting at $\pm2\sigma$, with $\sigma\approx25, 32$ and 42 $\mu$Jy beam$^{-1}$. The red cross indicates the VLA 1.4 GHz position (Morrison et al. 2010). All windows have identical axis. \label{fig:1}}
\end{figure*}

\section{Observations}

\subsection{BzK-21000}

Observations of the $^{12}$CO($J$=1--0) emission line ($\nu_\mathrm{rest}=115.2712$ GHz) of this galaxy were made simultaneously while observing the CO(2--1) emission line in the bright  $z=4.05$ SMG, GN20 (VLA project: AC974). Description of these observations are presented in detail in \citet{Hodge2012}. In summary, observations were carried out in D and B array configuration during 2010 March-April (D-configuration) and 2011 February-April (B-configuration). At the redshift of BzK-21000, $z_{\rm CO}=1.5213$ \citep{Daddi2010a}, the CO(1--0) line is shifted to 45.7184 GHz. The WIDAR correlator was configured with two spectral windows (SPWs) of 64 channels and 2 MHz per channel resolution. The SPWs were centered at 45.592 and 45.720 GHz for a total bandwidth of 246 MHz. The difference in total bandwidth with respect to $2x128=256$ MHz was due to overlapping channels, with the goal to improve the bandpass through the rejection of edge channels. The tuning offset with respect to the CO(1--0) sky frequency for BzK-21000 was chosen to cover the CO lines for 4 galaxies with the same narrow bandpass \citep[for details see ][]{Hodge2012, Hodge2013}. Similarly, the pointing center was selected to cover all 4 galaxies, such that the flux scaling multiplicative factor needed for primary beam correction at the location of BzK-21000 was 1.33.

The nearby quasar J1302+5748 was used for gain and pointing calibration and the source J1331+3030 (3C286) served as flux and bandpass calibrator. The data was calibrated using the Astronomical Image Processing System \citep[AIPS; ][]{Greisen2003} and the Common Astronomy Software Application (CASA) package \citep{McMullin2007}. Time ranges with poor visibilities as well as edge channels where the bandpass deteriorated (at each edge) were flagged. The data was imaged using the CLEAN algorithm in CASA. All images were accordingly primary beam corrected. We used a {\it briggs} weighting scheme, cleaning down to $2\sigma$ in a tight box ($\sim2.5''$ in size) around our target and setting the {\it robust} parameter equals to 0.8. We tried with other weighting schemes, however, we found that most of the emission was lost when using lower values for the {\it robust} parameter. Similarly, the images were dominated by large scale structures when using {\it robust}$>1.0$ or natural weighting. These structures were due to the combination of data of two difference configurations. Our selection is thus a compromise between obtaining good image quality without losing spatial information. A summary of the observations is presented in Table \ref{tab:0}.

\subsection{BzK-16000 and BzK-4171}
\begin{table*}
\centering
\caption{Observed properties of the high-redshift star-forming galaxies\label{tab:1}}
\begin{tabular}{lccccccccc}
\hline
Source & $z_\mathrm{CO}$ $^{\rm a}$& $v_{\rm FWHM}$ $^{\rm b}$ & $I_\mathrm{CO}$ $^{\rm c}$& $L'_\mathrm{CO}$ $^{\rm d}$     & $M_\mathrm{gas}$ $^{\rm e}$ & SFR $^{\rm f}$&    SFE $^{\rm g}$           &    $R_{21}$ $^{\rm h}$ &    $R_{31}$ $^{\rm h}$\\
              &                                  & (km s$^{-1}$)         &  (Jy km s$^{-1}$)                 & ($l_0$) & ($m_0$)       &   (M$_\odot$ yr$^{-1}$)      &    ($l_1$)       &                     \\
\hline\hline
BzK-21000 & 1.5204 (3)  &   $435\pm98$   & $0.20\pm0.05$        & $2.39\pm0.60$    & $0.86\pm0.21$  &   $210$   & 88 &  $0.80^{+0.27}_{-0.16}$ & $0.39^{+0.24}_{-0.14}$ \\
BzK-4171   &  1.464 (3)    &  $410\pm200$ & $0.22\pm0.07$         & $2.45\pm0.78$   & $0.88\pm0.28$ & $95$        & 38 &  $0.77^{+0.35}_{-0.18}$ &  -- \\
BzK-16000 & 1.524 (2)    & $217\pm80$     & $0.20\pm0.06$        & $2.40\pm0.72$    & $0.86\pm0.26$  &   $74$      & 30 &  $0.58^{+0.25}_{-0.13}$ & -- \\
BX610        & 2.2105 (3)   & $240\pm70$     & $0.18\pm0.04$        & $4.28\pm1.08$    & $1.54\pm0.39$  &   $212$       & 50 & --                                  	 & $0.58^{+0.21}_{-0.13}$ \\
\hline
\end{tabular}
\begin{flushleft}
\begin{footnotesize}
\noindent {\bf Notes:}
\noindent 
$^a$ CO(1--0) redshift. The parenthesis corresponds to the uncertainty associated to the last digit in this measurement; 
$^{\rm b}$ Measured CO line full-width at half maximum (FWHM); 
$^c$ Spatially and velocity integrated CO line flux $I_\mathrm{CO}=S_\mathrm{CO} dv$; 
$^d$ CO luminosity in units $l_0=10^{10}$ K km s$^{-1}$ pc$^2$; 
$^e$ Gas mass obtained using $\alpha_\mathrm{CO}=3.6\ M_\odot$ (K km s$^{-1}$ pc$^2)^{-1}$, in units $m_0=10^{11} (\alpha_{\rm CO}/3.6)\ M_\odot$; 
$^f$ Star formation rates (SFRs) derived from {\it Herschel} IR observations \citep{Magdis2012}, and UV/24$\mu$m measurements for BX610 \citep{Tacconi2013}. The associated uncertainties amount to $\sim$30\%. We assume a conversion SFR $=10^{-10} L_{\rm IR}$ and a \citet{Chabrier2003} Initial Mass function.
$^g$ Star formation efficiency derived from the CO(1--0) observations, defined as SFE $=L_\mathrm{IR}/L'_\mathrm{CO}$ in units of $l_1=L_\odot$ (K km s$^{-1}$ pc$^{2})^{-1}$; 
$^h$ Observed brightness temperature line ratios between the CO(3--2) and CO(2--1) with respect to the CO(1--0) line, where $R_{21}=T_{21}/T_{10}=I_{21}/I_{10}\times(\nu_{10}/\nu_{21})^2$ and similarly for $R_{31}$.
\end{footnotesize}
\end{flushleft}
\end{table*} 

\begin{figure*}
\centering
\includegraphics[scale=0.33]{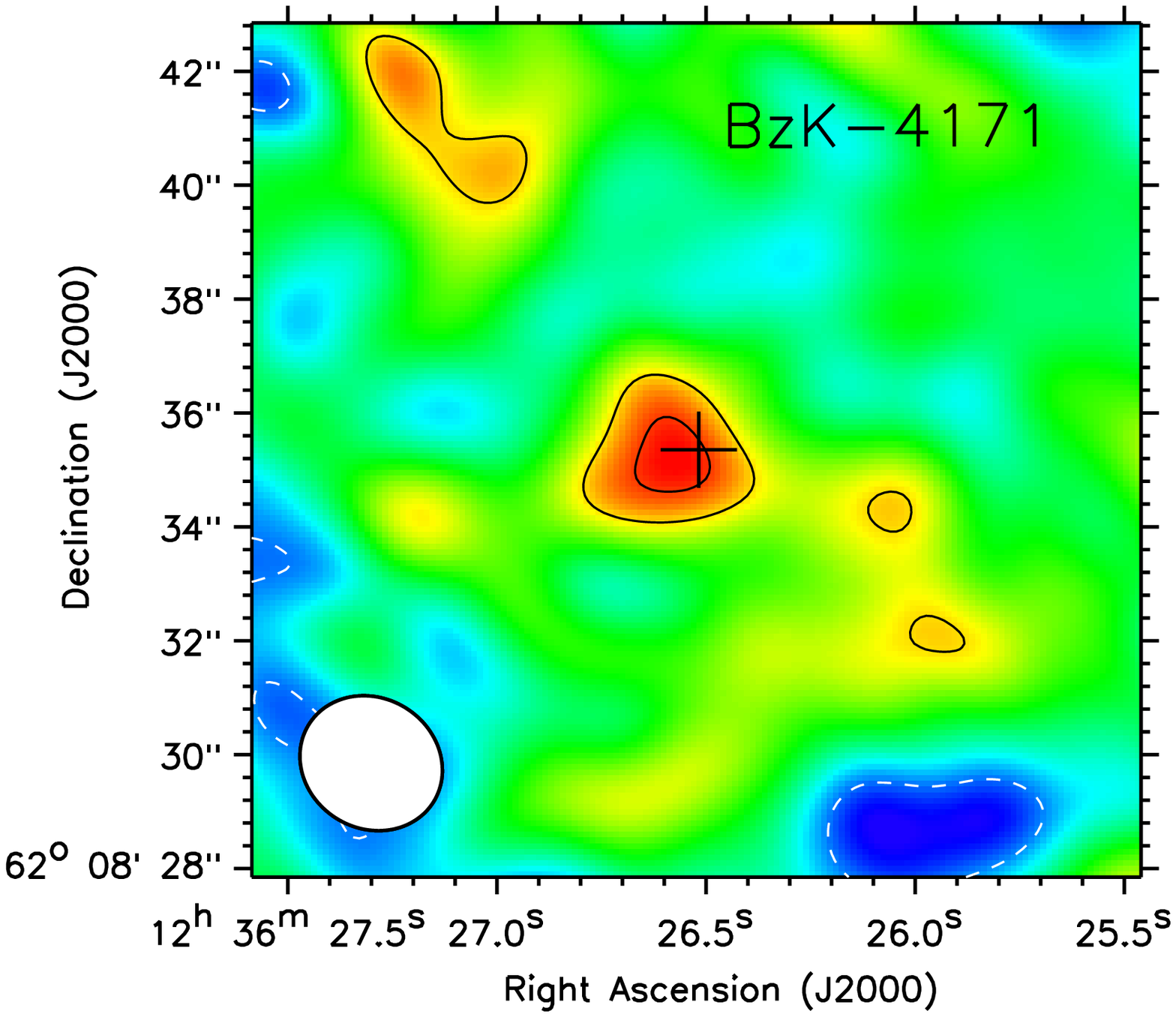} \hspace{4mm}
\includegraphics[scale=0.33]{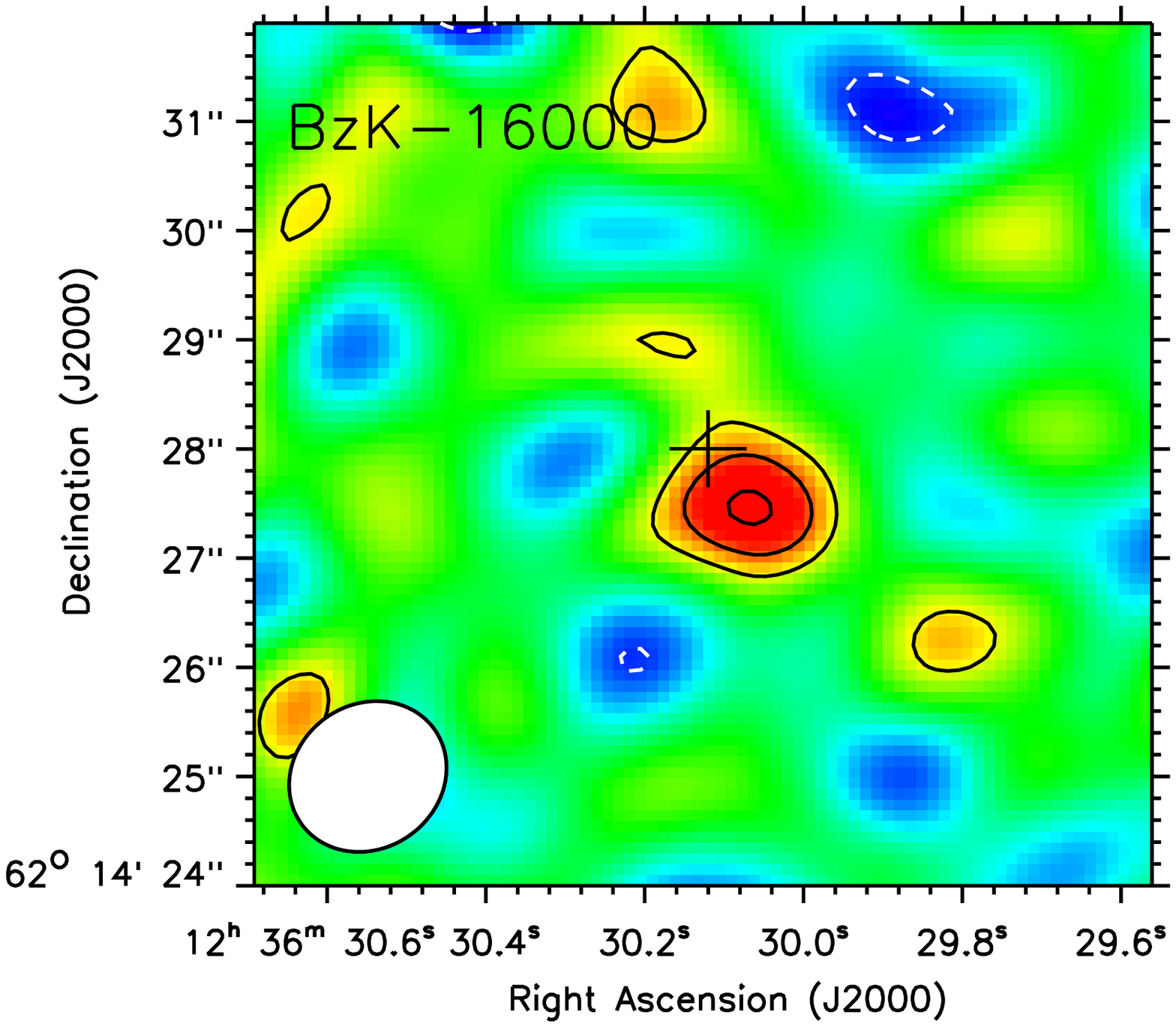}\hspace{4mm}
\includegraphics[scale=0.33]{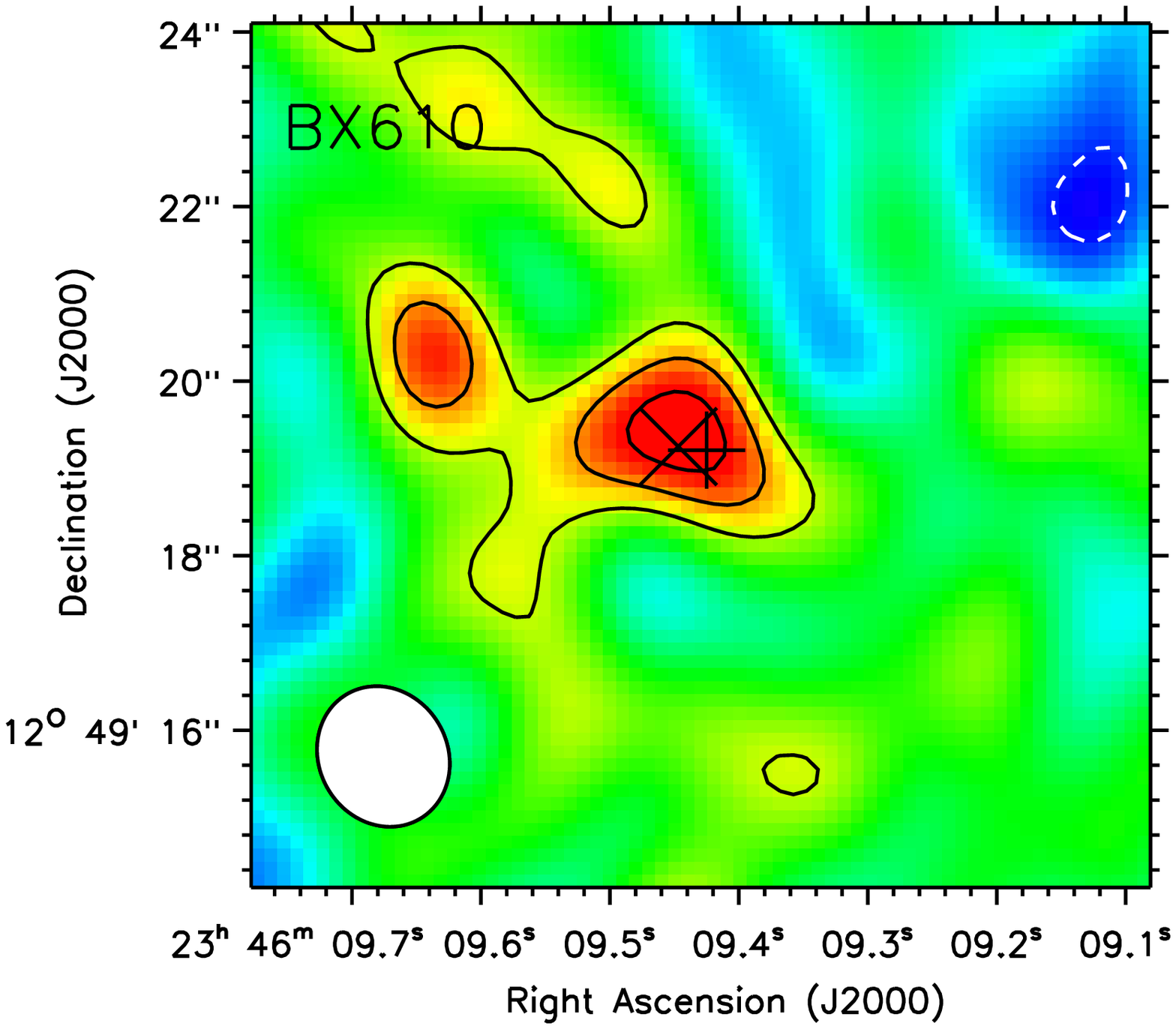}
\caption{VLA CO(1--0) maps averaged over 500, 200 and 300 km s$^{-1}$ for BzK-4171, BzK-16000 and  BX-610. Contour levels are given in steps of $\pm1\sigma$, starting at $\pm2\sigma$, with $\sigma\approx76, 100$ and  39 $\mu$Jy beam$^{-1}$ for each galaxy, respectively. The $+$ symbols indicate the VLA 1.4 GHz position for the BzK galaxies \citep{Morrison2010}, and the optical position for BX610 \citep{Erb2006,ForsterSchreiber2009}. The $\times$ symbol shows the CO(3--2) position for BX610 \citep{Tacconi2013}. \label{fig:2}}
\end{figure*}

CO(1--0) observations of BzK-4171 and BzK-16000 were obtained in the D, and D plus C array configurations, respectively. D-array observations of both sources were performed during 2010 March and 2010 April (VLA project: AC975). C-array observations  of BzK-16000 were done during 2012 April (VLA project 12A-264). At the redshifts $z_\mathrm{CO(2-1)}=1.4652$ and $z_{\rm CO(2-1)}=1.5249$ \citep{Daddi2010a}, the CO(1--0) line is shifted to 46.7594 GHz and 45.6537 GHz for BzK-4171 and BzK-16000, respectively. 

The calibrators and correlator configuration used for the D-array data were the same as those used to observe BzK-21000. For the BzK-4171, the SPWs were centered at 46.740 and 46.822 GHz, while for BzK-16000, these were centered at 45.647 and 45.765 GHz, for a total bandwidth of 228 MHz. Similarly to the case of BzK-21000, the difference in bandwidth with respect to the nominal 256 MHz is due to overlapping channels to improve the bandpass. For the C-array data, the observations were performed with the WIDAR correlator but using its expanded capabilities. In this case, two basebands (AC and BD) of 8 contiguous SPWs each were used. Baseband AC covered the SPWs 0-7 while baseband BD covered SPWs 8-15. Each SPW was set to have 64 channels and 2 MHz per channel resolution. Both basebands were configured to overlap, with an offset of 44 channels with respect to each other and making an effective bandwidth of $\sim1.112$ GHz. With this setup, the SPWs 5 and 12 covered the redshifted CO(1--0) line from BzK-16000. Hereafter, we focus on the data obtained in these SPWs. Both D- and C-array visibility datasets were concatenated and imaged accordingly. Data calibration and flagging were done with AIPS, and imaging was performed using CASA. After trying several weighting schemes, we chose {\it Briggs} weighting with a {\it robust} parameter of $1.0$. This approach minimized negative noise peaks in the image when averaging channels across the CO line, while retaining good sensitivity. In the case of BzK-4171, we applied a further taper of $1.7''$ to increase the sensitivity toward extended sources (although the actual rms of the map is higher). We used a box around the source of $\sim5''$ in size, cleaning down to $2\sigma$. The obtained sensitivities and beam sizes are summarized in Table \ref{tab:0}.

\subsection{BX610}
%\begin{figure*}
%\centering
%\includegraphics[scale=0.33]{bzk21000_spec_circle_freq_res8MHz_v2.ps}\hspace{4mm}
%\includegraphics[scale=0.33]{bzk4171_spec_peak_freq_res80kms.ps}\\
%\vspace{8mm}
%\includegraphics[scale=0.33]{bzk16000_spec_peak_freq_res80kms.ps}\hspace{4mm}
%\includegraphics[scale=0.33]{bx610_imcleanof_spec_peak_freq_res60kms.ps}
%\vspace{8mm}
%\caption{CO(1--0) spectra obtained for our targets. In the case of BzK-21000, we show the spectra obtained in a circular $2''$ aperture enclosing most of the CO emission. The spectral resolution is $\sim50$ km s$^{-1}$ for BzK-21000, $\sim60$ km s$^{-1}$ for BX610, and $\sim80$ km s$^{-1}$ for BzK-4171 and BzK-16000. The red arrows indicate the frequency of the CO(1--0) line based on the previously determined CO(2--1) redshift \citep{Daddi2010a}. For BX610, the red arrow shows the frequency of the CO(1--0) line using the optical redshift \citep{Erb2006, ForsterSchreiber2009, ForsterSchreiber2011}. \label{fig:3}}
%\end{figure*}

\begin{figure*}
\centering
\includegraphics[scale=0.26]{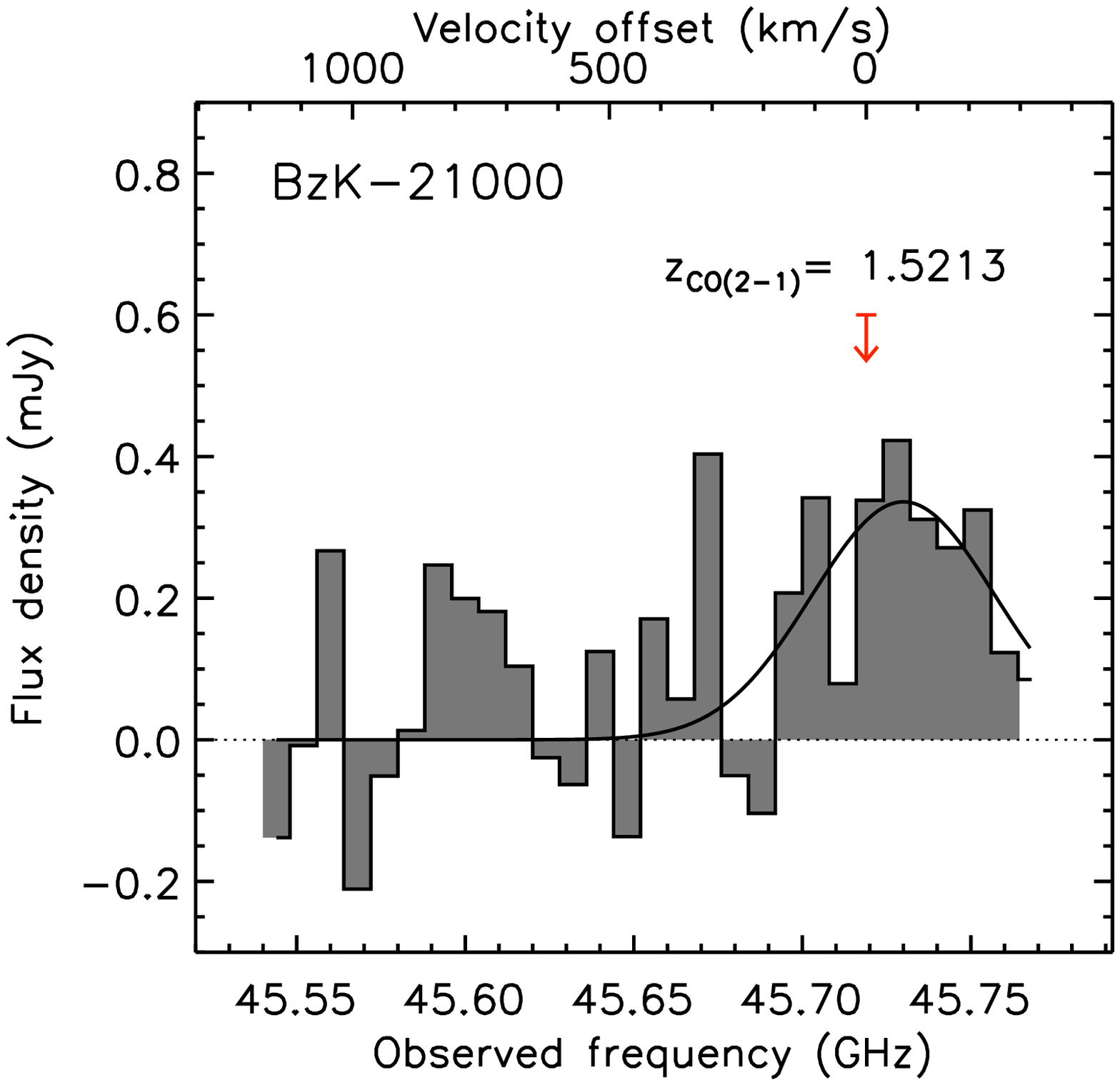}\hspace{1mm}
\includegraphics[scale=0.26]{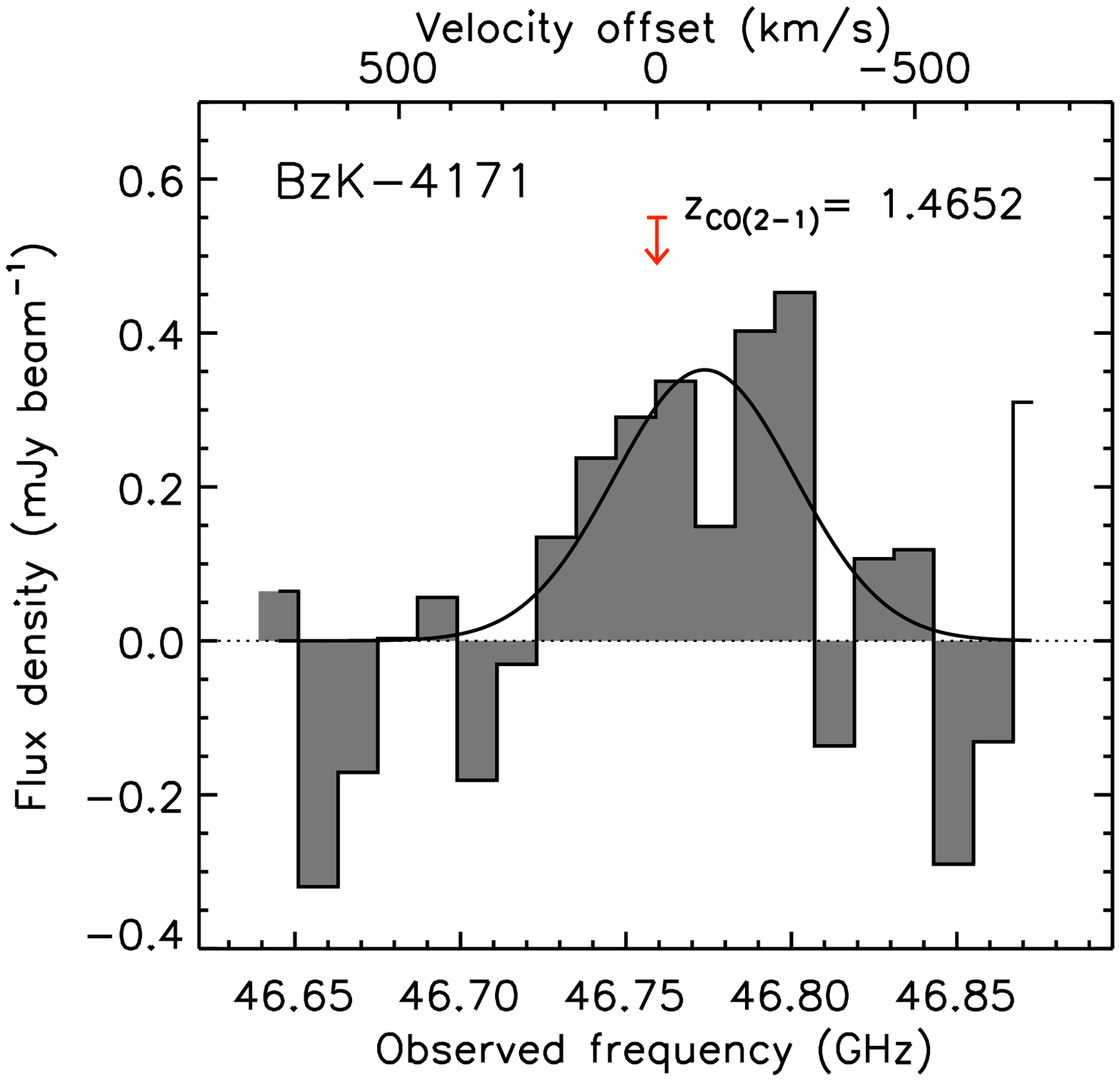}\hspace{1mm}
\includegraphics[scale=0.26]{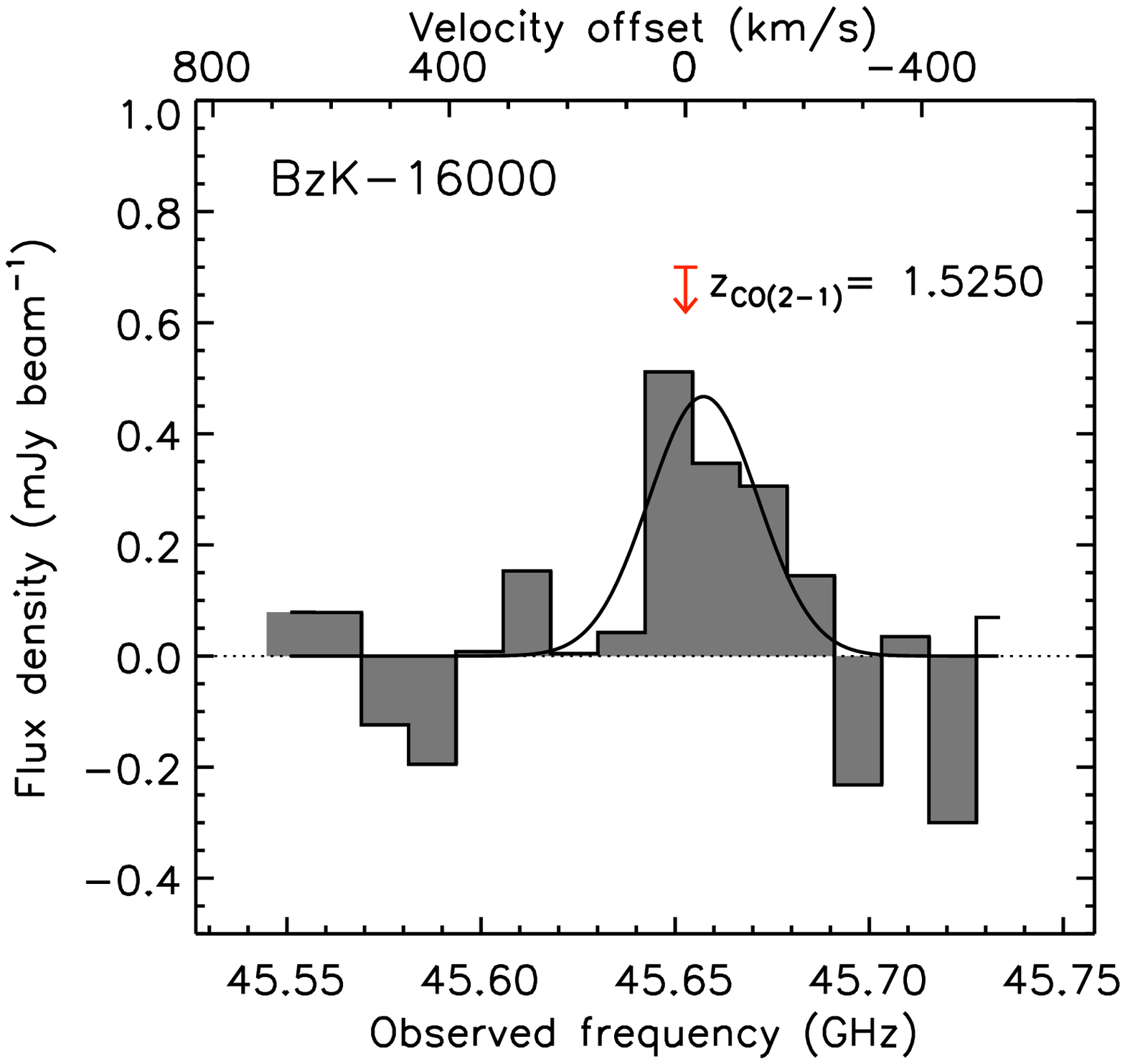}\hspace{1mm}
\includegraphics[scale=0.26]{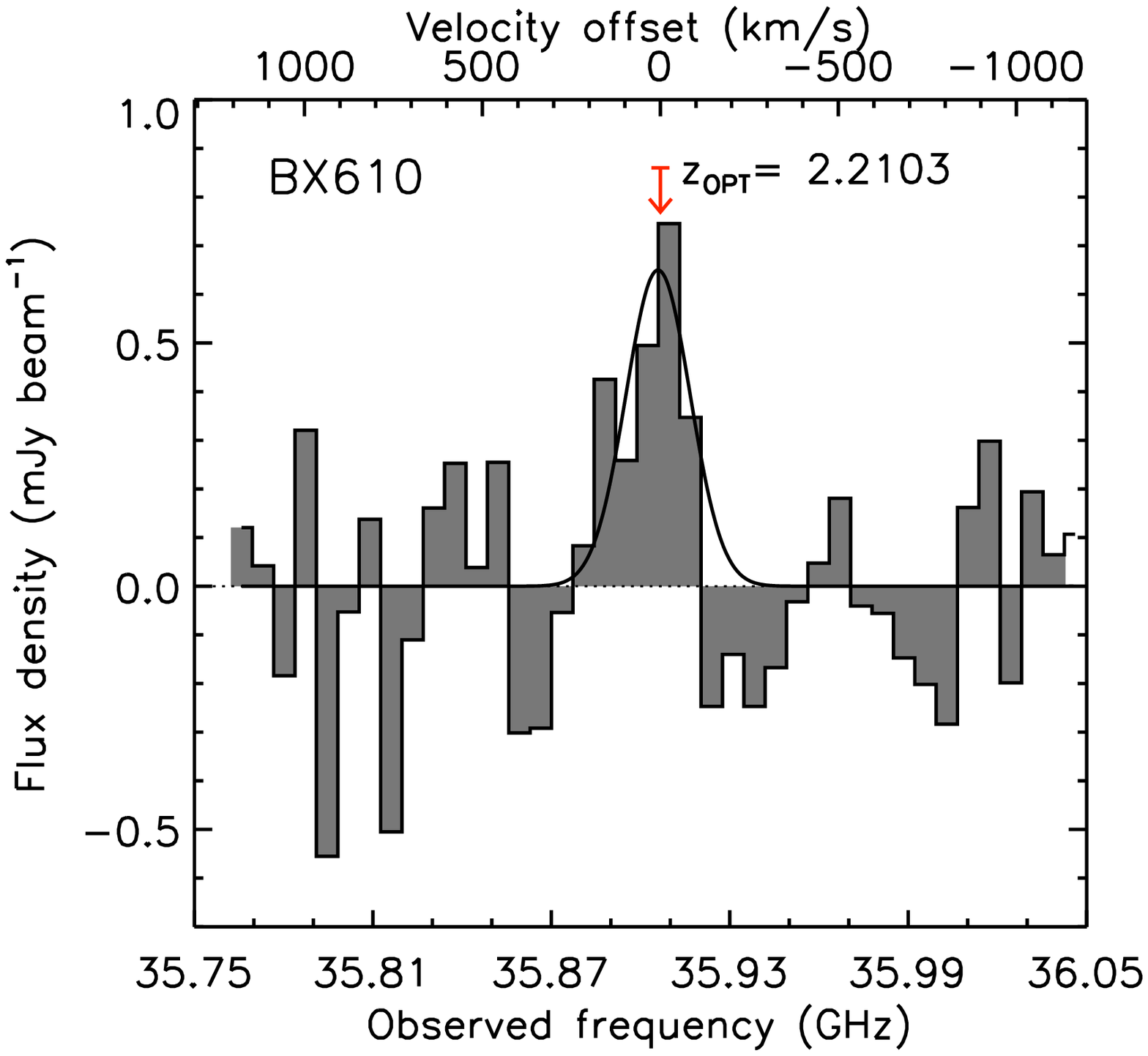}
\caption{CO(1--0) spectra obtained for our targets. In the case of BzK-21000, we show the spectra obtained in a circular $2''$ aperture enclosing most of the CO emission. The spectral resolution is $\sim50$ km s$^{-1}$ for BzK-21000, $\sim60$ km s$^{-1}$ for BX610, and $\sim80$ km s$^{-1}$ for BzK-4171 and BzK-16000. The red arrows indicate the frequency of the CO(1--0) line based on the previously determined CO(2--1) redshift \citep{Daddi2010a}. For BX610, the red arrow shows the frequency of the CO(1--0) line using the optical redshift \citep{Erb2006, ForsterSchreiber2009, ForsterSchreiber2011}. \label{fig:3}}
\end{figure*}

Observations of the CO(1--0) emission line in BX610 were performed from October 2011 to November 2011 in D-array configuration (VLA project: 11B-112), and pointed to the optical position listed in \citet{Erb2006}. These observations were performed with the correlator configured in a similar fashion as the BzK-16000 C-array observations. We used two basebands formed by 8 SPWs each, and the SPWs set with 64 channels and 2 MHz per channel resolution. Both basebands overlap, with an offset of 7 channels. With this setup, the SPWs 3 and 11 covered the redshifted CO(1--0) line. At the optical redshift $2.2103$ the CO(1--0) line is shifted to 35.9067 GHz. The source J2346+0930 was used for gain and pointing calibration. The sources J0137+3309 (3C48) and J2253+1608 (3C454.3) were used for flux and bandpass calibration, respectively. Data calibration, flagging and imaging was done using the CASA package. The data were mapped using {\it natural} weighting and cleaning down to $\sim2\sigma$ around a tight box around the source. The final sensitivity and beam are listed in Table \ref{tab:0}.

\section{Results and Analysis}

CO(1--0) line emission is detected in all galaxies in the sample. Figures \ref{fig:1}, \ref{fig:2} and \ref{fig:3} show the obtained CO maps and respective spectra. Table \ref{tab:1} summarizes the results. 

BzK-21000 is the only source where the resolution of the observations can probe smaller spatial scales. Here, the CO emission is resolved into several ``clumps''.  Given the complex structure of the CO emission, and in order to measure the full line intensity across the galaxy, we created a new CO {\it moment-0} map, tapering at a lower resolution ($1.1''$; Fig. \ref{fig:1}). While the source is still marginally resolved at this resolution, most of the internal structures are smoothed over. In the case of BzK-16000, BzK-4171 and BX-610, the emission is unresolved at the resolution and significance of the observations.

We derive total integrated line intensities from the velocity-integrated maps, by fitting Gaussians that were constrained by the position and size of the sources from the previous higher$-J$ CO observations with the Plateau de Bure Interferometer \citep[PdBI; ][]{Daddi2010a, Tacconi2013}. The CO intensities are listed in Table \ref{tab:1}. In the case of BzK-21000, this value includes a 12\% correction that takes into account the part of the line that falls outside the VLA tuning. The poor significance of the detections leads to small spatial offsets between the optical and CO(1--0) positions which might affect the derived CO fluxes. In the sample of BzK galaxies, the previous CO(1--0) observations are within $1\sigma$ of the new CO intensities (Table \ref{tab:1}) with the exception of BzK-21000, where the difference is at the $1.4\sigma$ level \citep{Aravena2010b}. This suggests that such offsets have only a minor effect in the final CO intensities. Also, as shown in Fig. \ref{fig:2}, the CO(1--0) peak position is in agreement with the CO(3--2) position for BX610.

% but they appear to show slightly different line widths, with $v_{\rm FWHM}=421\pm136$ km s$^{-1}$ and $v_{\rm FWHM}=565\pm256$ km s$^{-1}$ for the north and south component, respectively. These line-widths are still consistent within the uncertainties with the line-width computed for the whole region
Figure \ref{fig:3} shows the CO(1--0) spectra obtained for all sources. In the case of BzK-21000, where the emission is clearly resolved into more than a single component, we plot the spatially integrated spectrum. Since we are not resolving the CO emission in the other objects at the significance of the detections, we show the spectra at the peak position for these cases. By fitting a single Gaussian to the spectra we find the CO redshifts and line-widths listed in Table \ref{tab:1}. The latter compares well with the line profiles shown by \citet{Daddi2010a} and \citet{Tacconi2013}, however, any difference should not significantly affect our estimate of the CO integrated intensity, as it corresponds to the integral along the spectra from the moment-0 map.

%The large bandwidth covered for the observations of BX610 allow us to place a limit to its continuum emission at rest-frame 115 GHz of 57$\mu$Jy ($3\sigma$).

\subsection{CO luminosities and gas masses}

We derive CO luminosities through $L'_\mathrm{CO}=3.25\times10^7 (1+z)^{-3} \nu_\mathrm{obs}^{-2} I_\mathrm{CO} D_\mathrm{L}^2$, where $D_\mathrm{L}$ is the luminosity distance at redshift $z$, in Mpc, $\nu_\mathrm{obs}$ is the observing frequency, in GHz, and $I_\mathrm{CO}$ is the line intensity in Jy km s$^{-1}$ \citep{Solomon1997}. Using the measured CO line intensities, we obtain the CO luminosities listed in Table \ref{tab:1}. 

To estimate the gas masses from CO measurements, a conversion factor between CO luminosity and gas mass, $\alpha_\mathrm{CO}$ is commonly used (in units $M_\odot$ [K km s$^{-1}$ pc$^2$]$^{-1}$) such that $M_{\rm gas} = \alpha_{\rm CO} L'_{\rm CO(1-0)}$. Local Ultra-Luminous Infrared Galaxies (ULIRGs) were found to be consistent with $\alpha_\mathrm{CO}=0.8$ \citep{Downes1998}, however, in the Milky Way this factor was found to be $4.36$ \citep[e.g., ][]{Bolatto2013}. The latter is consistent with recent findings that suggest $\alpha_\mathrm{CO}=3.6$ in $z\sim1.5$ disk galaxies \citep{Daddi2010a}, however its actual value appears to be dependent on metallicity and highly uncertain \citep[e.g., ][]{Genzel2012}. Individual values of $\alpha_\mathrm{CO}$ for the BzK galaxies range from $\sim2.5-4.0$ with typical uncertainties of $\pm1.4$ \citep{Daddi2010a}.

Since most of our targets correspond to the ones used by Daddi et al. to derive this conversion factor, we adopt $\alpha_\mathrm{CO}=3.6$ throughout. The molecular gas masses obtained thereby are listed in Table \ref{tab:1}. 

\subsection{Line ratios}

Assuming the CO lines can be approximated in the Rayleigh-Jeans (RJ) limit, we can estimate the ratio between the CO(2--1) and CO(1--0) as $R_{21}=T_{21}/T_{10}=I_{21}/I_{10}\times(\nu_{10}/\nu_{21})^2$, where $T_{21}$ and $T_{10}$ are the brightness temperatures of the CO(2--1) and CO(1--0) lines, $I$ correspond to the CO line intensities, and $\nu_{21}$ and $\nu_{10}$ are the rest-frame CO line frequencies. Similarly, for the ratio between CO(3--2) and CO(1--0) we have $R_{31}=T_{32}/T_{10}=I_{32}/I_{10}\times(\nu_{10}/\nu_{32})^2$. The corresponding line ratios thereby obtained are listed in Table \ref{tab:1}. 

For local thermodynamic equilibrium (LTE) of the molecular gas, we would expect $R_{J1}=1$. However, in all cases we find that the ratios $R_{21}$ and $R_{31}$ are below unity, indicating that the gas is already sub-thermally excited at the CO(2--1) line. In the cases of BzK-4171 and BzK-16000, the new observations confirm within the uncertainties the previously derived values for $R_{21}$ \citep{Dannerbauer2009,Aravena2010b}. In the case of BzK-21000, the larger CO(1--0) intensity measured, compared to the previous measurements \citep{Aravena2010b}, imply a lower $R_{21}$ value yet still consistent with both lines being close to LTE. From these, we find a noise weighted average ratio between the 3 galaxies $R_{21}=0.70\pm0.16$. For the two galaxies with CO(3--2) measurements, we derive a noise weighted average $R_{31}=0.50\pm0.29$, where the quoted uncertainty includes the errors in the line ratio measurements ($\delta=+0.16$) and the scatter around the average ratio ($\delta=+0.13$). 

%\citet{Riechers2010} measured the $R_{31}$ line ratio in two lensed Lyman break galaxies (LBGs) at $z\sim2.7$, each having a ratio of $\sim0.75$. These galaxies, however, were found consistent with starbursting activity being not consistent with ``main-sequence'' galaxies, differential lensing could artificially enhance the fluxes measured for the CO(3--2) line if they arise from more compact regions compared to CO(1--0).

\subsection{Star-formation efficiencies}
%\begin{figure}[t]
%\centering
%\includegraphics[scale=0.72]{bzk21000_redblueoptical_mod.ps}
%\caption{CO(1--0) maps of the blue and redshifted emission, represented by the blue and red contours, overlaid on the HST $z$-band image for BzK-21000. Contour levels start at $+2\sigma$ in steps of $+1\sigma$. \label{fig:bluered}}
%\end{figure}
The SFR per unit mass of molecular gas, or the inverse of the gas depletion timescale, is usually taken as a measure of the star formation efficiency, SFE $={\rm SFR}/M_{\rm gas}$. Given the large uncertainties in the calibrations when obtaining SFRs from IR luminosities and molecular gas masses from CO luminosities (the $\alpha_{\rm CO}$ factor), a proxy for the SFE has also been defined as the ratio between the IR luminosity and the CO(1--0) luminosity, with SFE $=L_\mathrm{IR}/L'_\mathrm{CO}$ in units of $L_\odot$ (K km s$^{-1}$ pc$^{2}$)$^{-1}$. The SFEs for our sample are listed in Table \ref{tab:1}. As an effect of the revision in the line ratios in the previous section, the SFEs appear to be slightly lower than previously reported for these galaxies, yet well within the range found for local spiral galaxies with SFEs ranging from 10 to 100 $L_\odot$ (K km s$^{-1}$ pc$^{2}$)$^{-1}$ \citep[e.g.][]{Leroy2008}. SFE values for ULIRGs and distant SMGs typically exceed 100 $L_\odot$ (K km s$^{-1}$ pc$^{2}$)$^{-1}$ \citep[e.g.,][]{Yao2003, Neri2003,Greve2005, Bouche2007}.

%To estimate the SFEs in our sample, we first compute IR luminosities for all galaxies based on the available SFRs \citep{Magdis2012, Tacconi2010, Tacconi2013}. We convert these SFRs into IR luminosities assuming a Chabrier initial mass function \citep[IMF; ][]{Chabrier2003}, using $L_\mathrm{IR} (L_\odot)=1\times10^{10}$ SFR ($M_\odot$ yr$^{-1}$). This selection of IMF is consistent with that used by both Daddi et al., Magdis et al. and Tacconi et al. studies. Uncertainties in the SFRs amount to $\sim35\%$ \citep{Tacconi2013}. Note these do not include the uncertainties on the assumed IMF, which may differ by a factor of a few ($\sim2$) in the IR-derived SFR. From this, we obtain the SFEs listed in Table \ref{tab:1}. 

%{\bf To estimate the SFEs in our sample, we use the IR luminosities available from far-IR observations \citep{Magdis2012} or from UV/24$\mu$m calibration \citep{Tacconi2010, Tacconi2013}. Throughout, we assume a Chabrier initial mass function \citep[IMF; ][]{Chabrier2003} with SFR ($M_\odot$ yr$^{-1}$) = $1\times 10^{-10} L_\mathrm{IR} (L_\odot)$. This selection of IMF is consistent with that used by both Daddi et al., Magdis et al. and Tacconi et al. studies. Typical uncertainties in the IR luminosities and SFRs amount to $\sim35\%$ \citep{Tacconi2013}. Note these do not include the uncertainties on the assumed IMF, which may differ by a factor of a few ($\sim2$) in the IR-derived SFR. From this, we obtain the SFEs listed in Table \ref{tab:1}. }

\section{Discussion}

\subsection{Gas excitation}
\begin{figure}
\centering
\includegraphics[scale=0.35]{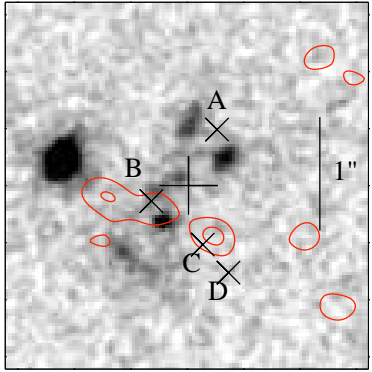}
\caption{HST ACS I band image toward BzK-21000. The contours represent the CO map tapered to $0.4''$ and averaged over 440 km s$^{-1}$ at levels of $+2, +3, +4\sigma$. The $\times$ symbols show the location of the candidate clumps identified in the data cube. The central cross represents the radio position of this galaxy. A kernel of 3 pixels has been used to smooth the HST image. The astrometry of the HST images is expected to be accurate to $<0.15''$ \citep{Daddi2007}. \label{fig:bzk2100_color}}
\end{figure}

Even though we only have two sources with $R_{31}$ values measured, and clearly more measurements are needed, our results support the typically assumed line ratio $R_{31}=0.5$. However, this also suggests that individual ratios can span a significant range from $\sim$ 0.4 to 0.6, and possibly larger given the uncertainties. Thus, while it is valid in a statistical sense to assume a constant value for this ratio, such an assumption may not necessarily be applicable to individual cases. This uncertainty can be considered to be relevant and at the same level of other involved sources of uncertainty such as low significance of CO detections, the $\alpha_{\rm CO}$ conversion factor, etc. 

% Indeed, our results suggest that assuming such ratio will add an uncertainty of $\sim50\%$ (based on the measured uncertainty in the average ratio) to the estimated CO(1--0) luminosity and gas mass. 
Our average $R_{31}$ value compares well with the average ratio $R_{31}=0.46\pm0.07$ found for three disk galaxies at $z\sim0.3$ \citep{Bauermeister2013}, which have SFRs of 40--60 M$_\odot$ yr$^{-1}$. Such low gas excitation has also been found in some SMGs at $z\sim2$, with an average value of $R_{31}\sim0.5-0.6$ \citep[][]{Harris2010,Ivison2011, Riechers2011a, Danielson2011, Bothwell2013, Spilker2014}. However, these line ratios might be the result of CO(1--0) being more spatially extended than higher-J CO lines with the line ratios being considerably higher in their centers. Distant quasar host galaxies (and radio galaxies) appear to be mostly thermalized at the CO(3--2) line \citep[e.g.,][]{Riechers2011c, Ivison2008, Ivison2012}, highlighting the role of AGN in molecular gas excitation in the host galaxy. This suggest that in disks and SMGs the gas excitation, and thus $R_{31}$, might be more influenced by star formation activity that it is in quasar host galaxies where the powerful AGN may play a more important role in gas heating.

%To investigate this, larger samples of galaxies, for which we know their nature (e.g. starbursts, quasars, disks), need to be observed in both cold and warm gas tracers, particularly in the CO(3--2) and CO(1--0) lines.

\subsection{Finding clumps in CO data: BzK-21000}

\begin{figure}
%\centering
\flushleft
%\vspace{6mm}
%\includegraphics[scale=0.4]{21000_WM1_spec_w2_ORI_flux_freq.ps}\hspace{12mm} %A
%\includegraphics[scale=0.4]{21000_WM2_spec_w1_ORI_flux_freq.ps}\hspace{12mm} %B
%\\ 
%\vspace{6mm}%D
%\includegraphics[scale=0.4]{21000_W23_spec_w1_ORI_flux_freq.ps}\hspace{12mm} %C
%\includegraphics[scale=0.4]{21000_W42_spec_w2_ORI_flux_freq.ps}\vspace{6mm} %D
\includegraphics[scale=0.5]{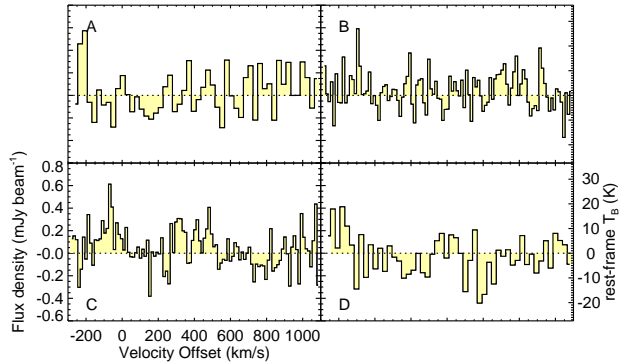}
\caption{CO line profiles measured at the location of the candidate clumps in the $0.18''$ resolution data cube. The right $y$-axis show the rest-frame brightness temperature obtained from the RJ approximation based on the flux density measured. The solid line represents a Gaussian fit to the CO line profiles. Profiles A and D are shown at 26 km s$^{-1}$ per channel resolution. Profiles B and C are at $13$ km s$^{-1}$ per channel resolution. \label{fig:clump_co}}
\end{figure}

The high resolution achieved in the observations of BzK-21000 enables us to look for bright compact CO clumps. While the map tapered to $1.1''$ resolution only reaches a significance of $\sim4.5\sigma$ for the integrated CO emission of the galaxy, compact clumps with larger brightness temperatures ($T_{\rm B}$) might be identified in the high resolution CO data cube.  

We searched for clumps using the AIPS task SERCH, which carries out a matched-filter analysis using a Gaussian kernel to convolve the data cube along the frequency axis with an expected input line width, and thereby searches for regions with a significance over a specified limit. This method has been used for source finding in a number of studies \citep[e.g.,][]{Hodge2012}. The expected size of the molecular clumps range from 0.3--1.0 kpc based on the size of star forming and stellar clumps from observations \citep[e.g.,][]{Swinbank2010,Jones2010,ForsterSchreiber2011} and numerical simulations \citep[e.g.,][]{Bournaud2007,Elmegreen2008,Dekel2009b}. We thus use a data cube in the native angular resolution of $0.18''$, or 1.5 kpc at $z=1.52$ (slightly larger than the expected clump size) and channel width of 52 km s$^{-1}$. We searched for clumps within a region of $3.0''\times3.0''$ centered at the VLA radio position of BzK-21000 \citep{Morrison2010}, utilizing Gaussian kernels in the range $\sim$ 50 to 200 km s$^{-1}$ in steps of 2 channels ($\sim50$ km s$^{-1}$) with a signal-to-noise cut at $S/N=4$. As a cross check, we independently inspected the data cube and manually searched for bright knots. We find that 4 line candidates can be identified in this way. Based on the number of negative peaks in the cube, however, we expect that 2 of these candidate clumps are false detections.

Figure \ref{fig:bzk2100_color} shows  the spatial location of each of the identified CO knots in the cube compared to the optical and  average CO images of BzK-21000. Figure \ref{fig:clump_co} shows the CO profiles for each of these candidates. Only one of the identified clump candidates (clump $B$) appears to coincide with an optical feature and with positive emission in the average CO map. In this case, however, the associated CO line profile appears to be the narrowest ($\sim20$ km s$^{-1}$). The disagreement in position between CO and optical clumps could indicate differential obscuration within the galaxy, where optical clumps represent less obscured regions poorer in molecular gas, however at the significance of the observations it is also possible that some of the CO candidates are false. Astrometric errors are unlikely given the accuracy of the HST images to $<0.15''$.

Using a Gaussian fit to the CO profiles, assuming a Rayleigh-Jeans approximation and accounting for redshift (factor $1+z$), we obtain brightness temperatures in the range $\sim10-35$ K (rest-frame) and line FWHM in the range $20-100$ km s$^{-1}$. Based on this, and the conservative assumption that $\alpha_\mathrm{CO}=3.6\ M_\odot$ (K km s$^{-1}$ pc$^2$)$^{-1}$, much larger than the value found for giant molecular clouds in our Galaxy \citep[$\alpha_\mathrm{CO}=0.6$; ][]{Oka1998}, we obtain gas masses in the range $(4.2-9.0)\times10^9\ M_\odot$. Together, these compact regions would make up to $<40\%$ of the total gas mass of BzK-21000, suggesting that a significant fraction of the molecular gas is diffuse or distributed in smaller and fainter clouds within this system. Using a lower conversion factor, more appropriate for giant molecular clouds, would lead to a significantly lower contribution from these CO clumps to the total gas mass in the galaxy. 

%{\bf We note that we search for clumps only as an obvious exercise in the highest spatial resolution CO data of this kind of unlensed object reported to date but care should be exercised when measuring physical properties for individual clump candidates.}

%To date, studies to isolate clumps in distant star-forming galaxies have relied on optical spectroscopic and continuum imaging tracing the ionized gas, a tracer of the star formation activity, and the stellar components of such galaxies \citep[e.g.;][]{ForsterSchreiber2011}. Such studies are able to locate individual clumps resolving the emission down to $\sim0.5-1.0$ kpc scales at the order of giant molecular complexes. While they have provided us with clear evidence that most of the high redshift disk galaxies fragment into massive regions, as predicted by some numerical simulations \citep[e.g.,][]{Bournaud2007,Elmegreen2008,Dekel2009b}, it is interesting to see how the molecular gas component, typically hidden to optical studies due to obscuration, of the ISM behaves at such scales. 

%Sub-arcsecond resolution measurements at radio/submillimeter wavelengths of the CO emission in high-redshift galaxies were possible in the past \citep[e.g.,][]{Riechers2008,Riechers2009, Swinbank2010, Swinbank2011, Freundlich2013},  but most of them were limited in bandwidth and velocity resolution thus hindering detailed clump searches. 

Few sub-arcsecond resolution studies have been able to search for CO clumps in high redshift galaxies \citep[e.g.,][]{Riechers2008,Riechers2009, Swinbank2010, Swinbank2011, Hodge2012,Freundlich2013}, mostly due to limitations in bandwidth, velocity resolution and sensitivity. Gravitational lensing has helped to resolve the emission in some systems. \citet{Swinbank2011} shows clear evidence for CO clumps in the bright gravitationally lensed disk galaxy SMM J2135--0102, finding clump sizes of 100--300 pc with velocity dispersions of $\sim$40--80 km s$^{-1}$. In this case, the molecular clumps constitute $\sim10-20\%$ the total gas mass of the system (for $\alpha_\mathrm{CO}=2$). Their results are similar to what is found more recently in the clump search toward the $z=4.05$ SMG GN20 \citep{Hodge2012}, and consistent with our limit for BzK-21000.

%Measurements of the brightness temperatures of CO peaks through velocity channels in the quasar host galaxy BRI 1335-0417 at $z=4.41$ show peak brightness temperatures in the range $29-61$ K within the central few kpc, comparable to the average kinetic gas and dust temperature in the host galaxy \citep{Riechers2008b}. More recently, 
%Recently, \citet{Hodge2012} performed a systematic search for clumps toward the $z=4.05$ SMG GN20, identifying 5 candidates with gas masses in the range $\sim(5-6)\times10^{9}\ M_\odot$ and brightness temperatures of $\sim16-31$ K. As pointed out by these authors \citep[see also ][]{Riechers2008b}, such temperatures are just in line with the galaxy dust temperature of $33\pm2$ K \citep{Magdis2011} and slightly lower than the high-excitation gas component with a kinetic temperature of $45$ K, suggesting that the clumps are close to being resolved at the spatial resolution of those observations. 

Independent of whether the candidate CO clumps are real or not in BzK-21000, it is clear that their brightness temperatures are below the average dust temperature in the galaxy of $34\pm2$ K \citep{Magdis2011}, implying that these gas clumps are either narrow in velocity ($\Delta v<25$ km s$^{-1}$) or have sizes that are smaller than the current spatial resolution ($<1.5$ kpc in size), or both.

\section{Conclusions}

We have presented detections of the CO(1--0) line emission in four massive star-forming galaxies in the redshift range $z=1.5-2.2$. Our observations allow us to confirm previous measurements of the brightness temperature line ratio between the CO(2--1) to CO(1--0) lines in three BzK galaxies.  We measure average line ratios of $R_{21}=0.70\pm0.16$ (based on the 3 BzK galaxies) and $R_{31}=0.50\pm0.29$ (based on two objects). These findings indicate that the molecular gas is, on-average, likely sub-thermally excited already in the CO(2--1). We find that the gas is sub-thermal in both cases at the CO(3--2) line, and support the widely assumed line ratio $R_{31}\sim0.5$ \citep{Tacconi2010, Tacconi2013, Genzel2013}. Since $R_{31}$ ranges from $\sim$ 0.4 to 0.6, we note that care must be exercised in assuming the average line ratio for individual cases.

Finally, we took advantage of the high angular resolution achieved for BzK-21000, and searched for gas clumps in the CO data cube. We found 4 clump candidates with estimated gas masses that can account for up to 40\% of the total molecular gas mass of this system. 

%We argue that while such searches are just at the limit achievable with current instrumentation, ALMA will be able to easily identify individual clumps in massive galaxies at high-redshift albeit in higher-J CO transitions. 

\section*{Acknowledgements}

MA and this work was co-funded under the Marie Curie Actions of the European Commission (FP7-COFUND). CC thanks the Kavli Institute for Cosmology for their hospitality. The National Radio Astronomy Observatory is a facility of the National Science Foundation (NSF), operated under cooperative agreement by Associated Universities Inc.

\bibliographystyle{mn2e}
\bibliography{cogalv2}

\begin{thebibliography}{62}
\expandafter\ifx\csname natexlab\endcsname\relax\def\natexlab#1{#1}\fi

\bibitem[{{Ao} {et~al}\mbox{.}(2008){Ao}, {Wei{\ss}}, {Downes}, {Walter},
  {Henkel}, \& {Menten}}]{Ao2008}
{Ao} Y., {Wei{\ss}} A., {Downes} D., {Walter} F., {Henkel} C., {Menten} K.~M.,
  2008, \aap, 491, 747

\bibitem[{{Aravena} {et~al}\mbox{.}(2010){Aravena}, {Carilli}, {Daddi}, {Wagg},
  {Walter}, {Riechers}, {Dannerbauer}, {Morrison}, {Stern}, \&
  {Krips}}]{Aravena2010b}
{Aravena} M. {et~al.}, 2010, \apj, 718, 177

\bibitem[{{Bauermeister} {et~al}\mbox{.}(2013){Bauermeister}, {Blitz},
  {Bolatto}, {Bureau}, {Teuben}, {Wong}, \& {Wright}}]{Bauermeister2013}
{Bauermeister} A., {Blitz} L., {Bolatto} A., {Bureau} M., {Teuben} P., {Wong}
  T., {Wright} M., 2013, \apj, 763, 64

\bibitem[{{Bolatto} {et~al}\mbox{.}(2013){Bolatto}, {Wolfire}, \&
  {Leroy}}]{Bolatto2013}
{Bolatto} A.~D., {Wolfire} M., {Leroy} A.~K., 2013, \araa, 51, 207

\bibitem[{{Bothwell} {et~al}\mbox{.}(2013){Bothwell}, {Smail}, {Chapman},
  {Genzel}, {Ivison}, {Tacconi}, {Alaghband-Zadeh}, {Bertoldi}, {Blain},
  {Casey}, {Cox}, {Greve}, {Lutz}, {Neri}, {Omont}, \&
  {Swinbank}}]{Bothwell2013}
{Bothwell} M.~S. {et~al.}, 2013, \mnras, 429, 3047

\bibitem[{{Bouch{\'e}} {et~al}\mbox{.}(2007){Bouch{\'e}}, {Cresci}, {Davies},
  {Eisenhauer}, {F{\"o}rster Schreiber}, {Genzel}, {Gillessen}, {Lehnert},
  {Lutz}, {Nesvadba}, {Shapiro}, {Sternberg}, {Tacconi}, {Verma}, {Cimatti},
  {Daddi}, {Renzini}, {Erb}, {Shapley}, \& {Steidel}}]{Bouche2007}
{Bouch{\'e}} N. {et~al.}, 2007, \apj, 671, 303

\bibitem[{{Bournaud} {et~al}\mbox{.}(2007){Bournaud}, {Elmegreen}, \&
  {Elmegreen}}]{Bournaud2007}
{Bournaud} F., {Elmegreen} B.~G., {Elmegreen} D.~M., 2007, \apj, 670, 237

\bibitem[{{Carilli} \& {Walter}(2013)}]{Carilli2013}
{Carilli} C.~L., {Walter} F., 2013, \araa, 51, 105

\bibitem[{{Chabrier}(2003)}]{Chabrier2003}
{Chabrier} G., 2003, \pasp, 115, 763

\bibitem[{{Daddi} {et~al}\mbox{.}(2010{\natexlab{a}}){Daddi}, {Bournaud},
  {Walter}, {Dannerbauer}, {Carilli}, {Dickinson}, {Elbaz}, {Morrison},
  {Riechers}, {Onodera}, {Salmi}, {Krips}, \& {Stern}}]{Daddi2010a}
{Daddi} E. {et~al.}, 2010{\natexlab{a}}, \apj, 713, 686

\bibitem[{{Daddi} {et~al}\mbox{.}(2008){Daddi}, {Dannerbauer}, {Elbaz},
  {Dickinson}, {Morrison}, {Stern}, \& {Ravindranath}}]{Daddi2008}
{Daddi} E., {Dannerbauer} H., {Elbaz} D., {Dickinson} M., {Morrison} G.,
  {Stern} D., {Ravindranath} S., 2008, \apjl, 673, L21

\bibitem[{{Daddi} {et~al}\mbox{.}(2005){Daddi}, {Dickinson}, {Chary}, {Pope},
  {Morrison}, {Alexander}, {Bauer}, {Brandt}, {Giavalisco}, {Ferguson}, {Lee},
  {Lehmer}, {Papovich}, \& {Renzini}}]{Daddi2005}
{Daddi} E. {et~al.}, 2005, \apjl, 631, L13

\bibitem[{{Daddi} {et~al}\mbox{.}(2007){Daddi}, {Dickinson}, {Morrison},
  {Chary}, {Cimatti}, {Elbaz}, {Frayer}, {Renzini}, {Pope}, {Alexander},
  {Bauer}, {Giavalisco}, {Huynh}, {Kurk}, \& {Mignoli}}]{Daddi2007}
{Daddi} E. {et~al.}, 2007, \apj, 670, 156

\bibitem[{{Daddi} {et~al}\mbox{.}(2010{\natexlab{b}}){Daddi}, {Elbaz},
  {Walter}, {Bournaud}, {Salmi}, {Carilli}, {Dannerbauer}, {Dickinson},
  {Monaco}, \& {Riechers}}]{Daddi2010b}
{Daddi} E. {et~al.}, 2010{\natexlab{b}}, \apjl, 714, L118

\bibitem[{{Danielson} {et~al}\mbox{.}(2011){Danielson}, {Swinbank}, {Smail},
  {Cox}, {Edge}, {Weiss}, {Harris}, {Baker}, {De Breuck}, {Geach}, {Ivison},
  {Krips}, {Lundgren}, {Longmore}, {Neri}, \& {Flaquer}}]{Danielson2011}
{Danielson} A.~L.~R. {et~al.}, 2011, \mnras, 410, 1687

\bibitem[{{Dannerbauer} {et~al}\mbox{.}(2009){Dannerbauer}, {Daddi},
  {Riechers}, {Walter}, {Carilli}, {Dickinson}, {Elbaz}, \&
  {Morrison}}]{Dannerbauer2009}
{Dannerbauer} H., {Daddi} E., {Riechers} D.~A., {Walter} F., {Carilli} C.~L.,
  {Dickinson} M., {Elbaz} D., {Morrison} G.~E., 2009, \apjl, 698, L178

\bibitem[{{Dekel} {et~al}\mbox{.}(2009){Dekel}, {Sari}, \&
  {Ceverino}}]{Dekel2009b}
{Dekel} A., {Sari} R., {Ceverino} D., 2009, \apj, 703, 785

\bibitem[{{Downes} \& {Solomon}(1998)}]{Downes1998}
{Downes} D., {Solomon} P.~M., 1998, \apj, 507, 615

\bibitem[{{Elmegreen} {et~al}\mbox{.}(2008){Elmegreen}, {Bournaud}, \&
  {Elmegreen}}]{Elmegreen2008}
{Elmegreen} B.~G., {Bournaud} F., {Elmegreen} D.~M., 2008, \apj, 688, 67

\bibitem[{{Erb} {et~al}\mbox{.}(2006){Erb}, {Steidel}, {Shapley}, {Pettini},
  {Reddy}, \& {Adelberger}}]{Erb2006}
{Erb} D.~K., {Steidel} C.~C., {Shapley} A.~E., {Pettini} M., {Reddy} N.~A.,
  {Adelberger} K.~L., 2006, \apj, 646, 107

\bibitem[{{F{\"o}rster Schreiber} {et~al}\mbox{.}(2009){F{\"o}rster Schreiber},
  {Genzel}, {Bouch{\'e}}, {Cresci}, {Davies}, {Buschkamp}, {Shapiro},
  {Tacconi}, {Hicks}, {Genel}, {Shapley}, {Erb}, {Steidel}, {Lutz},
  {Eisenhauer}, {Gillessen}, {Sternberg}, {Renzini}, {Cimatti}, {Daddi},
  {Kurk}, {Lilly}, {Kong}, {Lehnert}, {Nesvadba}, {Verma}, {McCracken},
  {Arimoto}, {Mignoli}, \& {Onodera}}]{ForsterSchreiber2009}
{F{\"o}rster Schreiber} N.~M. {et~al.}, 2009, \apj, 706, 1364

\bibitem[{{F{\"o}rster Schreiber} {et~al}\mbox{.}(2011){F{\"o}rster Schreiber},
  {Shapley}, {Genzel}, {Bouch{\'e}}, {Cresci}, {Davies}, {Erb}, {Genel},
  {Lutz}, {Newman}, {Shapiro}, {Steidel}, {Sternberg}, \&
  {Tacconi}}]{ForsterSchreiber2011}
{F{\"o}rster Schreiber} N.~M. {et~al.}, 2011, \apj, 739, 45

\bibitem[{{Freundlich} {et~al}\mbox{.}(2013){Freundlich}, {Combes}, {Tacconi},
  {Cooper}, {Genzel}, {Neri}, {Bolatto}, {Bournaud}, {Burkert}, {Cox}, {Davis},
  {F{\"o}rster Schreiber}, {Garcia-Burillo}, {Gracia-Carpio}, {Lutz}, {Naab},
  {Newman}, {Sternberg}, \& {Weiner}}]{Freundlich2013}
{Freundlich} J. {et~al.}, 2013, \aap, 553, A130

\bibitem[{{Genzel} {et~al}\mbox{.}(2008){Genzel}, {Burkert}, {Bouch{\'e}},
  {Cresci}, {F{\"o}rster Schreiber}, {Shapley}, {Shapiro}, {Tacconi},
  {Buschkamp}, {Cimatti}, {Daddi}, {Davies}, {Eisenhauer}, {Erb}, {Genel},
  {Gerhard}, {Hicks}, {Lutz}, {Naab}, {Ott}, {Rabien}, {Renzini}, {Steidel},
  {Sternberg}, \& {Lilly}}]{Genzel2008}
{Genzel} R. {et~al.}, 2008, \apj, 687, 59

\bibitem[{{Genzel} {et~al}\mbox{.}(2012){Genzel}, {Tacconi}, {Combes},
  {Bolatto}, {Neri}, {Sternberg}, {Cooper}, {Bouch{\'e}}, {Bournaud},
  {Burkert}, {Comerford}, {Cox}, {Davis}, {F{\"o}rster Schreiber},
  {Garcia-Burillo}, {Gracia-Carpio}, {Lutz}, {Naab}, {Newman}, {Saintonge},
  {Shapiro}, {Shapley}, \& {Weiner}}]{Genzel2012}
{Genzel} R. {et~al.}, 2012, \apj, 746, 69

\bibitem[{{Genzel} {et~al}\mbox{.}(2006){Genzel}, {Tacconi}, {Eisenhauer},
  {F{\"o}rster Schreiber}, {Cimatti}, {Daddi}, {Bouch{\'e}}, {Davies},
  {Lehnert}, {Lutz}, {Nesvadba}, {Verma}, {Abuter}, {Shapiro}, {Sternberg},
  {Renzini}, {Kong}, {Arimoto}, \& {Mignoli}}]{Genzel2006}
{Genzel} R. {et~al.}, 2006, \nat, 442, 786

\bibitem[{{Genzel} {et~al}\mbox{.}(2010){Genzel}, {Tacconi}, {Gracia-Carpio},
  {Sternberg}, {Cooper}, {Shapiro}, {Bolatto}, {Bouch{\'e}}, {Bournaud},
  {Burkert}, {Combes}, {Comerford}, {Cox}, {Davis}, {Schreiber},
  {Garcia-Burillo}, {Lutz}, {Naab}, {Neri}, {Omont}, {Shapley}, \&
  {Weiner}}]{Genzel2010}
{Genzel} R. {et~al.}, 2010, \mnras, 407, 2091

\bibitem[{{Genzel} {et~al}\mbox{.}(2013){Genzel}, {Tacconi}, {Kurk}, {Wuyts},
  {Combes}, {Freundlich}, {Bolatto}, {Cooper}, {Neri}, {Nordon}, {Bournaud},
  {Burkert}, {Comerford}, {Cox}, {Davis}, {F{\"o}rster Schreiber},
  {Garc{\'{\i}}a-Burillo}, {Gracia-Carpio}, {Lutz}, {Naab}, {Newman},
  {Saintonge}, {Shapiro Griffin}, {Shapley}, {Sternberg}, \&
  {Weiner}}]{Genzel2013}
{Genzel} R. {et~al.}, 2013, \apj, 773, 68

\bibitem[{{Greisen}(2003)}]{Greisen2003}
{Greisen} E.~W., 2003, Information Handling in Astronomy - Historical Vistas,
  285, 109

\bibitem[{{Greve} {et~al}\mbox{.}(2005){Greve}, {Bertoldi}, {Smail}, {Neri},
  {Chapman}, {Blain}, {Ivison}, {Genzel}, {Omont}, {Cox}, {Tacconi}, \&
  {Kneib}}]{Greve2005}
{Greve} T.~R. {et~al.}, 2005, \mnras, 359, 1165

\bibitem[{{Harris} {et~al}\mbox{.}(2010){Harris}, {Baker}, {Zonak}, {Sharon},
  {Genzel}, {Rauch}, {Watts}, \& {Creager}}]{Harris2010}
{Harris} A.~I., {Baker} A.~J., {Zonak} S.~G., {Sharon} C.~E., {Genzel} R.,
  {Rauch} K., {Watts} G., {Creager} R., 2010, \apj, 723, 1139

\bibitem[{{Hodge} {et~al}\mbox{.}(2013){Hodge}, {Carilli}, {Walter}, {Daddi},
  \& {Riechers}}]{Hodge2013}
{Hodge} J.~A., {Carilli} C.~L., {Walter} F., {Daddi} E., {Riechers} D., 2013,
  \apj, 776, 22

\bibitem[{{Hodge} {et~al}\mbox{.}(2012){Hodge}, {Carilli}, {Walter}, {de Blok},
  {Riechers}, {Daddi}, \& {Lentati}}]{Hodge2012}
{Hodge} J.~A., {Carilli} C.~L., {Walter} F., {de Blok} W.~J.~G., {Riechers} D.,
  {Daddi} E., {Lentati} L., 2012, \apj, 760, 11

\bibitem[{{Ivison} {et~al}\mbox{.}(2008){Ivison}, {Morrison}, {Biggs}, {Smail},
  {Willner}, {Gurwell}, {Greve}, {Stevens}, \& {Ashby}}]{Ivison2008}
{Ivison} R.~J. {et~al.}, 2008, \mnras, 390, 1117

\bibitem[{{Ivison} {et~al}\mbox{.}(2011){Ivison}, {Papadopoulos}, {Smail},
  {Greve}, {Thomson}, {Xilouris}, \& {Chapman}}]{Ivison2011}
{Ivison} R.~J., {Papadopoulos} P.~P., {Smail} I., {Greve} T.~R., {Thomson}
  A.~P., {Xilouris} E.~M., {Chapman} S.~C., 2011, \mnras, 412, 1913

\bibitem[{{Ivison} {et~al}\mbox{.}(2012){Ivison}, {Smail}, {Amblard},
  {Arumugam}, {De Breuck}, {Emonts}, {Feain}, {Greve}, {Haas}, {Ibar},
  {Jarvis}, {Kov{\'a}cs}, {Lehnert}, {Nesvadba}, {R{\"o}ttgering}, {Seymour},
  \& {Wylezalek}}]{Ivison2012}
{Ivison} R.~J. {et~al.}, 2012, \mnras, 425, 1320

\bibitem[{{Jones} {et~al}\mbox{.}(2010){Jones}, {Swinbank}, {Ellis}, {Richard},
  \& {Stark}}]{Jones2010}
{Jones} T.~A., {Swinbank} A.~M., {Ellis} R.~S., {Richard} J., {Stark} D.~P.,
  2010, \mnras, 404, 1247

\bibitem[{{Komatsu} {et~al}\mbox{.}(2011){Komatsu}, {Smith}, {Dunkley},
  {Bennett}, {Gold}, {Hinshaw}, {Jarosik}, {Larson}, {Nolta}, {Page},
  {Spergel}, {Halpern}, {Hill}, {Kogut}, {Limon}, {Meyer}, {Odegard}, {Tucker},
  {Weiland}, {Wollack}, \& {Wright}}]{Komatsu2011}
{Komatsu} E. {et~al.}, 2011, \apjs, 192, 18

\bibitem[{{Leroy} {et~al}\mbox{.}(2008){Leroy}, {Walter}, {Brinks}, {Bigiel},
  {de Blok}, {Madore}, \& {Thornley}}]{Leroy2008}
{Leroy} A.~K., {Walter} F., {Brinks} E., {Bigiel} F., {de Blok} W.~J.~G.,
  {Madore} B., {Thornley} M.~D., 2008, \aj, 136, 2782

\bibitem[{{Magdis} {et~al}\mbox{.}(2012){Magdis}, {Daddi}, {B{\'e}thermin},
  {Sargent}, {Elbaz}, {Pannella}, {Dickinson}, {Dannerbauer}, {da Cunha},
  {Walter}, {Rigopoulou}, {Charmandaris}, {Hwang}, \&
  {Kartaltepe}}]{Magdis2012}
{Magdis} G.~E. {et~al.}, 2012, \apj, 760, 6

\bibitem[{{Magdis} {et~al}\mbox{.}(2011){Magdis}, {Daddi}, {Elbaz}, {Sargent},
  {Dickinson}, {Dannerbauer}, {Aussel}, {Walter}, {Hwang}, {Charmandaris},
  {Hodge}, {Riechers}, {Rigopoulou}, {Carilli}, {Pannella}, {Mullaney},
  {Leiton}, \& {Scott}}]{Magdis2011}
{Magdis} G.~E. {et~al.}, 2011, \apjl, 740, L15

\bibitem[{{Magnelli} {et~al}\mbox{.}(2012){Magnelli}, {Saintonge}, {Lutz},
  {Tacconi}, {Berta}, {Bournaud}, {Charmandaris}, {Dannerbauer}, {Elbaz},
  {F{\"o}rster-Schreiber}, {Graci{\'a}-Carpio}, {Ivison}, {Maiolino}, {Nordon},
  {Popesso}, {Rodighiero}, {Santini}, \& {Wuyts}}]{Magnelli2012}
{Magnelli} B. {et~al.}, 2012, \aap, 548, A22

\bibitem[{{McMullin} {et~al}\mbox{.}(2007){McMullin}, {Waters}, {Schiebel},
  {Young}, \& {Golap}}]{McMullin2007}
{McMullin} J.~P., {Waters} B., {Schiebel} D., {Young} W., {Golap} K., 2007, in
  Astronomical Society of the Pacific Conference Series, Vol. 376, Astronomical
  Data Analysis Software and Systems XVI, {Shaw} R.~A., {Hill} F., {Bell}
  D.~J., eds., p. 127

\bibitem[{{Morrison} {et~al}\mbox{.}(2010){Morrison}, {Owen}, {Dickinson},
  {Ivison}, \& {Ibar}}]{Morrison2010}
{Morrison} G.~E., {Owen} F.~N., {Dickinson} M., {Ivison} R.~J., {Ibar} E.,
  2010, \apjs, 188, 178

\bibitem[{{Neri} {et~al}\mbox{.}(2003){Neri}, {Genzel}, {Ivison}, {Bertoldi},
  {Blain}, {Chapman}, {Cox}, {Greve}, {Omont}, \& {Frayer}}]{Neri2003}
{Neri} R. {et~al.}, 2003, \apjl, 597, L113

\bibitem[{{Oka} {et~al}\mbox{.}(1998){Oka}, {Hasegawa}, {Hayashi}, {Handa}, \&
  {Sakamoto}}]{Oka1998}
{Oka} T., {Hasegawa} T., {Hayashi} M., {Handa} T., {Sakamoto} S., 1998, \apj,
  493, 730

\bibitem[{{Papadopoulos} \& {Ivison}(2002)}]{Papadopoulos2002}
{Papadopoulos} P.~P., {Ivison} R.~J., 2002, \apjl, 564, L9

\bibitem[{{Riechers} {et~al}\mbox{.}(2011{\natexlab{a}}){Riechers}, {Carilli},
  {Maddalena}, {Hodge}, {Harris}, {Baker}, {Walter}, {Wagg}, {Vanden Bout},
  {Wei{\ss}}, \& {Sharon}}]{Riechers2011c}
{Riechers} D.~A. {et~al.}, 2011{\natexlab{a}}, \apjl, 739, L32

\bibitem[{{Riechers} {et~al}\mbox{.}(2011{\natexlab{b}}){Riechers}, {Carilli},
  {Walter}, {Weiss}, {Wagg}, {Bertoldi}, {Downes}, {Henkel}, \&
  {Hodge}}]{Riechers2011a}
{Riechers} D.~A. {et~al.}, 2011{\natexlab{b}}, \apjl, 733, L11

\bibitem[{{Riechers} {et~al}\mbox{.}(2011{\natexlab{c}}){Riechers}, {Hodge},
  {Walter}, {Carilli}, \& {Bertoldi}}]{Riechers2011b}
{Riechers} D.~A., {Hodge} J., {Walter} F., {Carilli} C.~L., {Bertoldi} F.,
  2011{\natexlab{c}}, \apjl, 739, L31

\bibitem[{{Riechers} {et~al}\mbox{.}(2009){Riechers}, {Walter}, {Bertoldi},
  {Carilli}, {Aravena}, {Neri}, {Cox}, {Wei{\ss}}, \& {Menten}}]{Riechers2009}
{Riechers} D.~A. {et~al.}, 2009, \apj, 703, 1338

\bibitem[{{Riechers} {et~al}\mbox{.}(2008){Riechers}, {Walter}, {Brewer},
  {Carilli}, {Lewis}, {Bertoldi}, \& {Cox}}]{Riechers2008}
{Riechers} D.~A., {Walter} F., {Brewer} B.~J., {Carilli} C.~L., {Lewis} G.~F.,
  {Bertoldi} F., {Cox} P., 2008, \apj, 686, 851

\bibitem[{{Riechers} {et~al}\mbox{.}(2006){Riechers}, {Walter}, {Carilli},
  {Knudsen}, {Lo}, {Benford}, {Staguhn}, {Hunter}, {Bertoldi}, {Henkel},
  {Menten}, {Weiss}, {Yun}, \& {Scoville}}]{Riechers2006}
{Riechers} D.~A. {et~al.}, 2006, \apj, 650, 604

\bibitem[{{Shapiro} {et~al}\mbox{.}(2008){Shapiro}, {Genzel}, {F{\"o}rster
  Schreiber}, {Tacconi}, {Bouch{\'e}}, {Cresci}, {Davies}, {Eisenhauer},
  {Johansson}, {Krajnovi{\'c}}, {Lutz}, {Naab}, {Arimoto}, {Arribas},
  {Cimatti}, {Colina}, {Daddi}, {Daigle}, {Erb}, {Hernandez}, {Kong},
  {Mignoli}, {Onodera}, {Renzini}, {Shapley}, \& {Steidel}}]{Shapiro2008}
{Shapiro} K.~L. {et~al.}, 2008, \apj, 682, 231

\bibitem[{{Solomon} {et~al}\mbox{.}(1997){Solomon}, {Downes}, {Radford}, \&
  {Barrett}}]{Solomon1997}
{Solomon} P.~M., {Downes} D., {Radford} S.~J.~E., {Barrett} J.~W., 1997, \apj,
  478, 144

\bibitem[{{Spilker} {et~al}\mbox{.}(2014){Spilker}, {Marrone}, {Aguirre},
  {Aravena}, {Ashby}, {Bethermin}, {Bradford}, {Bothwell}, {Brodwin},
  {Carlstrom}, {Chapman}, {Crawford}, {de Breuck}, {Fassnacht}, {Gonzalez},
  {Greve}, {Gullberg}, {Hezaveh}, {Holzapfel}, {Husband}, {Ma}, {Malkan},
  {Murphy}, {Reichardt}, {Rotermund}, {Stalder}, {Stark}, {Strandet}, {Vieira},
  {Weiss}, \& {Welikala}}]{Spilker2014}
{Spilker} J.~S. {et~al.}, 2014, arXiv:1403.1667

\bibitem[{{Swinbank} {et~al}\mbox{.}(2011){Swinbank}, {Papadopoulos}, {Cox},
  {Krips}, {Ivison}, {Smail}, {Thomson}, {Neri}, {Richard}, \&
  {Ebeling}}]{Swinbank2011}
{Swinbank} A.~M. {et~al.}, 2011, \apj, 742, 11

\bibitem[{{Swinbank} {et~al}\mbox{.}(2010){Swinbank}, {Smail}, {Longmore},
  {Harris}, {Baker}, {De Breuck}, {Richard}, {Edge}, {Ivison}, {Blundell},
  {Coppin}, {Cox}, {Gurwell}, {Hainline}, {Krips}, {Lundgren}, {Neri}, {Siana},
  {Siringo}, {Stark}, {Wilner}, \& {Younger}}]{Swinbank2010}
{Swinbank} A.~M. {et~al.}, 2010, \nat, 464, 733

\bibitem[{{Tacconi} {et~al}\mbox{.}(2010){Tacconi}, {Genzel}, {Neri}, {Cox},
  {Cooper}, {Shapiro}, {Bolatto}, {Bouch{\'e}}, {Bournaud}, {Burkert},
  {Combes}, {Comerford}, {Davis}, {Schreiber}, {Garcia-Burillo},
  {Gracia-Carpio}, {Lutz}, {Naab}, {Omont}, {Shapley}, {Sternberg}, \&
  {Weiner}}]{Tacconi2010}
{Tacconi} L.~J. {et~al.}, 2010, \nat, 463, 781

\bibitem[{{Tacconi} {et~al}\mbox{.}(2013){Tacconi}, {Neri}, {Genzel}, {Combes},
  {Bolatto}, {Cooper}, {Wuyts}, {Bournaud}, {Burkert}, {Comerford}, {Cox},
  {Davis}, {F{\"o}rster Schreiber}, {Garc{\'{\i}}a-Burillo}, {Gracia-Carpio},
  {Lutz}, {Naab}, {Newman}, {Omont}, {Saintonge}, {Shapiro Griffin}, {Shapley},
  {Sternberg}, \& {Weiner}}]{Tacconi2013}
{Tacconi} L.~J. {et~al.}, 2013, \apj, 768, 74

\bibitem[{{Wei{\ss}} {et~al}\mbox{.}(2007){Wei{\ss}}, {Downes}, {Neri},
  {Walter}, {Henkel}, {Wilner}, {Wagg}, \& {Wiklind}}]{Weiss2007}
{Wei{\ss}} A., {Downes} D., {Neri} R., {Walter} F., {Henkel} C., {Wilner}
  D.~J., {Wagg} J., {Wiklind} T., 2007, \aap, 467, 955

\bibitem[{{Yao} {et~al}\mbox{.}(2003){Yao}, {Seaquist}, {Kuno}, \&
  {Dunne}}]{Yao2003}
{Yao} L., {Seaquist} E.~R., {Kuno} N., {Dunne} L., 2003, \apj, 588, 771

\end{thebibliography}

%\bsp

\label{lastpage}

\end{document}